# Optomagnetic composite medium with conducting nanoelements

L. V. Panina, A. N. Grigorenko, and D. P. Makhnovskiy

*Department of Communication, Electronic and Electrical Engineering,*
*University of Plymouth, Drake Circus, Plymouth, Devon PL4 8AA, United Kingdom.*

**Abstract** A new type of metal-dielectric composites has been proposed that is characterised by a resonance-like behaviour of the effective permeability $\mu_{eff}$ in the infrared and visible spectral ranges. This material can be referred to as optomagnetic medium. It consists of conducting inclusions in the shape of non-closed contours or pairs of parallel sticks with length of 50-100 nm embedded in dielectric matrix. The analytical formalism developed is based on solving the scattering problem for considered inclusions with impedance boundary condition, which yields the current and charge distributions within the inclusions. The magnetic properties originated by induced currents are enhanced by localised plasmon modes, which make an inclusion resonate at a much lower frequency than that of the half-wavelength requirement at microwaves. It implies that microstructure can be made on a scale much less than the wavelength and the effective permeability is a valid concept. The presence of the effective magnetic permeability and its resonant properties lead to novel optical effects and open new possible applications. In particular, the condition for Brewster's angle becomes different resulting in reflectionless normal incidence from air (vacuum) if the effective permeability and permittivity are the same. The resonant behaviour of the effective permeability of the proposed optomagnetic medium could be used for creation of optical polarizes, filters, phase shifters, selective lenses.

## I. Introduction

Metal-dielectric composites, in which small metal particles are embedded into dielectric host, present an exciting area of study. The overall electric, magnetic and optical properties are not governed by the behaviour of the raw materials. Vast literature exists on this topic. In the limit of high metal concentration, the percolation across the connected clusters results in critical dielectric[1] and magnetic[2] responses, strong local field fluctuations[3] and enhancement of transport[4,5] and optical[6,7] nonlinearities. In the limit of diluted composites, individual metal inclusions contribute to the effective electromagnetic properties; however, small metallic scatters may show completely different behaviour as compared with bulk metals. In both cases, the effective permittivity $\varepsilon_{eff}$ and permeability $\mu_{eff}$ can be tuned to values not easily possible in natural materials.

Recent advances in microfabrications make possible creation of composite materials with constituents of different forms and sizes down to nanoscales.[8,9] This offers a way to engineer various dielectric and magnetic metamaterials, since the effective parameters $\varepsilon_{eff}$ and $\mu_{eff}$ are determined by microstructure. Composites containing rings, helix or $\Omega$-particles exhibit resonance-like behaviour of both the permittivity and permeability in overlapping frequency bands,[10,11] which is quite unusual in nature. In a medium of three-dimensional array of intersecting wires the propagation modes have a dispersion characteristic similar to that in a neutral plasma with negative $\varepsilon_{eff}$ below the plasma resonance somewhere in the gigahertz range.[12,13] It was further shown that the composites built of two-dimensional arrays of split copper rings[14] and wires have a range of frequencies over which both the permittivity and permeability are negative in microwaves.[15]

These materials have generated a considerable interest as they offer a possibility to realise a negative index of refraction, $n$. Many surprising effects are possible in these so-called left-handed materials theoretically predicted by Veselago,[16] which would be of great importance for communication and electronics. These include reversed Doppler and Cherenkov effects, reversed Snell's angle, which could result in lenses without limitations on the resolution by wavelength.[17] So far, the concept of negative refraction has been predicted and proven at microwave frequencies. In the experiment on deflection of a beam of microwave radiation by a prism made of wire-and-ring material negative refraction angles were found,[18] which correspond to the negative index $n$ appropriate to Snell's law. The transmission spectra measured in these materials also confirms the concept of negative $n$.[15]

An immediate question is whether left-handed materials can be realised at optical frequencies. (We exclude from consideration photonic crystals where it is difficult to assign an effective equivalent $n$ and where the phenomenon of negative refraction has been recently predicted near negative group velocity bands.[19]) On one hand, negative dielectric constant is natural for metals below the plasma resonance that falls above visible frequencies. For example, silver would be a good choice to have a negative permittivity at optical frequencies since the resistive part is very small. On the other hand, there are quite rigid limitations existing with respect to the permeability at high frequencies. There is a widespread believe that the concept of permeability has no physical meaning at optical frequencies and onward, as was proven for atomic magnetism (see, for example, Ref. 20). *The aim of this paper is to elucidate the implications related to high-frequency magnetic properties and to demonstrate that metal-dielectric composites with nano-inclusions can have a considerable magnetic activity at optical frequencies.*

We consider metal-dielectric composites with two types of inclusions forming current contours: two-wire contour and a single ring with a gap (see Fig. 1). The mathematical formalism developed is based on a modified antenna theory, which provides a bridge between the microwave methods using distributed parameters and the optical description based on plane waves and surface plasmons. We show that the magnetic properties at optical frequencies can be generated by localised plasmon modes. The theory predicts resonance-like behaviour of the effective permeability over certain frequency range in the infrared and visible parts of spectrum in such composites. The negative values of $\mu_{eff}$ are possible past the resonance; however, high volume fractions are needed to realise $\mu_{eff} < 0$ since the magnetic dipole interaction strongly reduces the resonance peaks. Nevertheless, other unusual optical effects are can be realised in the presence of some $\mu_{eff}$ noticeably different from unity (which do not require negative $\mu_{eff}$). In particularly, the condition under which there is no reflected waves (Brewster's angle) changes. In the case $\mu_{eff} = \varepsilon_{eff}$ a normal incidence from air (vacuum) gives no reflection. This effect known at microwaves could be useful for optical filters and isolators.

The phenomenon of non-trivial permeability at optical frequencies can be named as optomagnetism (and the area of study is then referred to as optomagnetics) in order to distinguish it from the magnetic field influence on light propagation known as magnetooptics. The proposed optomagnetic medium is foreseen to find a variety of applications in optics and optoelectronics. The resonance properties of the effective permeability at optical frequencies can be exploited for the production of tuneable narrow bandwidth optical filters (either on reflection or on absorption). The presence of permeability may be used in optoelectronic interferometers to measure phase shifts induced by magnetic fields and is likely to result in a new generation of all optical sensors. The dependence of permeability dispersion upon carrier concentration inside the inclusion and external parameters could be used for creation of tuneable optical elements.

Natural ferromagnetic behaviour tails off completely at gigaherz frequencies. In the case of ferrite materials with few magnetic sub-lattices, there exists so-called exchange modes of magnetic excitations with eigen frequencies in the infrared, but their intensity is very small.

Then, only the electron movement under the Lorentz force would contribute to $\mu_{eff}$ at optical frequencies. In this case, there appears a problem of using a concept of permeability in the averaged macroscopic Maxwell equations. The situation is different for composite materials containing nano-sized metal inclusions. The effective permeability is obtained by averaging the magnetic moments of closed currents in metal inclusions and can have quite high values as a result of the resonance interaction of the incident electromagnetic wave with plasmons confined inside the inclusion (localised plasmon modes). The resonant frequency is lowered considerably allowing the resonance to occur at wavelength of the exciting light much larger than the inclusion size (at microwaves, there would be a half-wavelength requirement for resonance) and the effective parameters are still a valid concept. Near a resonance, the currents in the inclusions are enhanced and large magnetic moments are generated. That may either enhance (paramagnetic effect) or oppose (diamagnetic effect) the incident field and the effective permeability exhibits a resonant behaviour. Considering a diluted system with a small volume concentration $p << 1$ of the current contours and calculating $\mu_{eff}$ by summation over independent magnetic moments give the dispersion of the effective permeability with $\mu_{eff} > 1$ below the resonance and $\mu_{eff}$ goes to negative values past the resonance. However, the interaction between the inclusions (taken into account within the effective medium theory[21,22]) decreases considerably the values of $\mu_{eff}$ near the resonance and for realistic concentrations ($p < 0.1$) always $\mu_{eff} > 0$. The considered system has also the effective permittivity $\varepsilon_{eff}$. For ring-inclusions, the value of $\varepsilon_{eff}$ is reduced due to geometry. For two-wire contours, the permittivity is essential for the light polarisation with the electric field along the wires (see Fig. 1(a)); however, the resonance frequency for $\varepsilon_{eff}$ is shifted towards higher frequencies with respect to that for $\mu_{eff}$, which makes it possible to realise the condition $\mu_{eff} \approx \varepsilon_{eff}$ in both cases.

The paper is organised as follows. Section II starts with the description of proposed optical effects due to existence of substantial $\mu_{eff}$. In Section III we discuss in more detail the limitations on the concept of permeability at high frequencies. Section IV formulates how to determine the effective parameters in the considered composites. In Section V, the mathematical formalism for calculation of the current distribution in the inclusions is given. Section VI presents the obtained results on $\mu_{eff}$ and $\varepsilon_{eff}$ and overall discussion. There are also two mathematical appendixes.

## II. Brewster's angle in the presence of permeability

Introducing effective magnetic permeability at optical frequencies may result in some unusual behaviour of light propagating in macroscopically heterogeneous composite media (with boundaries). Here we consider the reflection and refraction of a polarised light at an interface between two dielectric media with non-trivial magnetic properties. The results of this Section are, in many respects, known (see, for example, Ref. 23), however, it is convenient to re-examine the conditions of reflection/refraction described by Fresnel's equations, since they are customarily analysed for nonmagnetic materials. In particular, we show that the condition under which there is no a reflected wave from a boundary (Brewster's angle) becomes different in the presence of permeability. The light is incident from medium "1" with the material parameters $\varepsilon_1, \mu_1$ towards medium "2" with $\varepsilon_2, \mu_2$. The quantities pertaining to the incident, reflected and transmitted waves are distinguished by the suffixes "i", "r" and "t", respectively, as shown in Fig. 2. In the case of the electric field **E** perpendicular to the plane of incidence (s-polarisation) the Fresnel's relationship between the field in the incident wave and that in the reflected wave is:[23]

$$\left(\frac{E_r}{E_i}\right)_{\perp} = \frac{(n_1/\mu_1)\cos\theta_i - (n_2/\mu_2)\cos\theta_t}{(n_1/\mu_1)\cos\theta_i + (n_2/\mu_2)\cos\theta_t}. \qquad (1)$$

Here, $\theta_{i,r,t}$ are the corresponding angles of incidence, reflection and refraction obeying usual Snell's equation:

$$\theta_i = \theta_r, \quad n_1 \sin\theta_i = n_2 \sin\theta_t \tag{2}$$

with the index of refraction $n = \sqrt{\varepsilon\,\mu}$. The condition $E_r = 0$, or

$$n_1\mu_2 \cos\theta_i = n_2\mu_1 \cos\theta_t, \tag{3}$$

gives Brewster's angle $\theta_b$. For non-magnetic media $\mu_1 = \mu_2 = 1$ equation (3) can hold only if $\varepsilon_1 = \varepsilon_2$ (no optical interface). Therefore, Brewster's angle is not observed for s-polarization in conventional optics. With $\mu_{1,2} \neq 1$ the absence of reflection and Brewster's angle can happen even for s-polarisation:

$$\tan\theta_b = \sqrt{\frac{\varepsilon_1\mu_2^2 - \varepsilon_2\mu_1\mu_2}{\varepsilon_2\mu_1\mu_2 - \varepsilon_1\mu_1^2}} \tag{4}$$

when $(\varepsilon_1\mu_2^2 - \varepsilon_2\mu_1\mu_2)/(\varepsilon_2\mu_1\mu_2 - \varepsilon_1\mu_1^2) \geq 0$.

Suppose that the optical properties are due only to the magnetic permeability ($\varepsilon_1 = \varepsilon_2 = 1, n_1 = \sqrt{\mu_1}$, $n_2 = \sqrt{\mu_2}$), then, the Brewster angle is given by:

$$\tan\theta_b = n_2 / n_1. \tag{5}$$

Equation (5) formally coincides with a usual equation known for *p*-polarised light (electrical field is in the plane of incidence). For incidence from air (vacuum) ($\varepsilon_1 = \mu_1 = 1$), equation (4) becomes:

$$\tan\theta_b = \sqrt{\frac{\mu_2^2 - n_2^2}{n_2^2 - 1}},$$

which can be satisfied either by $\mu_2^2 \geq n_2^2 > 1$ ($|\mu_2| \geq |\varepsilon_2|$ and $|n_2| > 1$) or by $\mu_2^2 \leq n_2^2 < 1$ ($|\mu_2| \leq |\varepsilon_2|$ and $|n_2| < 1$).

For the $p$-polarisation case, the Brewster angle can be found from (4) by interchanging $\varepsilon$ and $\mu$. This is because the boundary conditions for the two cases are symmetrical with respect to **E** and **H**. Then, (4) becomes:

$$\tan\theta_b = \sqrt{\frac{\mu_1\varepsilon_2^2 - \mu_2\varepsilon_1\varepsilon_2}{\mu_2\varepsilon_1\varepsilon_2 - \mu_1\varepsilon_1^2}} \tag{6}$$

and Brewster's angle exists for p-polarization when $(\mu_1\varepsilon_2^2 - \mu_2\varepsilon_1\varepsilon_2)/(\mu_2\varepsilon_1\varepsilon_2 - \mu_1\varepsilon_1^2) \geq 0$. For $\mu_1 = \mu_2 = 1$ equation (6) gives a standard form for the Brewster angle: $\tan\theta_b = n_2 / n_1$ (compare with (5)). In the case of incidence from air ($\varepsilon_1 = 1$ and $\mu_1 = 1$) equation (6) reduces to:

$$\tan\theta_b = \sqrt{\frac{\varepsilon_2^2 - n_2^2}{n_2^2 - 1}}\,. \tag{7}$$

From (7) it is clear that Brewster's angle in this polarization is realised either under the condition $\varepsilon_2^2 \geq n_2^2 > 1$ ($|\varepsilon_2| \geq |\mu_2|$ and $|n_2| > 1$) or $\varepsilon_2^2 \leq n_2^2 < 1$ ($|\varepsilon_2| \leq |\mu_2|$ and $|n_2| < 1$), which are opposite to those for s-polarisation.

It is worth mentioning that for both polarisations there is no reflection at normal incidence under the condition $\varepsilon_1 / \mu_1 = \varepsilon_2 / \mu_2$, as it follows from (4) and (6). This result is well known for microwaves, representing the condition of the impedance matching since the ratio $\varepsilon / \mu$ is related to the wave impedance.[23] It also means that an arbitrary polarised light will not be reflected from the interface of air (vacuum) and a medium with optical constants $|\varepsilon_2| = |\mu_2|$. Thus, there may be an optical analogy of the impedance matching.

In a view of these conditions of reflection, optomagnetic materials may demonstrate new interesting phenomena at optical frequencies, suitable for applications in optical filters, phase-shifters and isolators.

## III. Limitations imposed on permeability at high frequencies

In this Section we analyse the restriction conditions for introducing high frequency permeability to the Maxwell equations. The magnetisation $\mathbf{M} = (\mathbf{B} - \mathbf{H})/4\pi$ appears in the Maxwell equations as a result of averaging the microscopic current density $\mathbf{J}_{mic}$:

$$\mathbf{curl}\,\mathbf{M} = \frac{<\mathbf{J}_{mic}>}{c}. \qquad (8)$$

Here <…> stands for a mean value of a microscopic quantity, $c$ is the velocity of light (Gaussian units are used throughout the paper). The physical meaning of magnetisation is the magnetic moment per unit volume. This comes from the possibility to rewrite the total magnetisation of the body in the form:

$$\int \mathbf{M}\,dV = \frac{1}{2c}\int (\mathbf{r} \times <\mathbf{J}_{mic}>) dV, \qquad (9)$$

where the integration is carried out inside the body. Equation (8) is correct for a static magnetic field. When the macroscopic fields depend on time, the establishment of the relationship between the mean value $<\mathbf{J}_{mic}>$ and other quantities is not straightforward. A general form of equation (8) is:[20]

$$\mathbf{curl}\mathbf{M} = \frac{<\mathbf{J}_{mic}>}{c} - \frac{1}{c}\frac{d\mathbf{P}}{dt}, \qquad (10)$$

where $\mathbf{P}$ is the polarisation vector. However, equation (10) is not consistent with (9). Therefore, the physical meaning of $\mathbf{M}$ at high frequencies depends on the possibility of neglecting the second term in the right part of equation (10):

$$c\,\mathbf{curl}\mathbf{M} >> \frac{d\mathbf{P}}{dt}. \qquad (11)$$

Let us suppose that the fields are induced by an electromagnetic wave of frequency $\omega$. Estimating $\mathbf{curl}\mathbf{M} \sim \chi H / l$, $dP/dt \sim \omega\alpha E$, where $\chi$ is the magnetic susceptibility, $\alpha$ is the electric polarisability, $l$ is the characteristic size of the system, and taking $E \sim H$ for the electromagnetic wave, (11) can be written as:

$$\chi >> \alpha l / \lambda \qquad (12)$$

where $\lambda$ is the wavelength. It is generally considered that at optical frequencies (and onward) inequality (12) cannot be satisfied and the concept of the permeability is meaningless. This is correct if the magnetic moment is associated with electron motion in the atom. Indeed, the relaxation times for any paramagnetic or ferromagnetic processes are considerably larger than optical periods. Then $\chi$ is due to electron movements under the Lorenz force, and can be estimated as $\chi \sim (v/c)^2$, where $v$ is the electron velocity in the atom. On the other hand, the optical frequencies are of the order of $v/b$ where $b$ is the atomic dimension. Then (12) reads as $l << b(v/c)$, which is not compatible with the requirement that the characteristic size of the system has to be much greater than the atomic dimension ($l >> b$).

The situation may be completely different for metal-dielectric composites. If the inclusion size is smaller than the wavelength (but larger than the atomic size) the effective magnetic and dielectric parameters can be introduced as a result of local field averaging. For a composite with metal grains of a few nanometres, the effective parameters including permeability become meaningful even for optical frequencies. In this case, the effective permeability is obtained by averaging the magnetic moments of induced currents in metal inclusions. Our analysis demonstrates that inequality (12) is satisfied in the composites considered over certain frequencies in the optical range.

## IV. Problem formulation

In metal-dielectric composites irradiated by high frequency electromagnetic field the magnetic properties are produced by contour currents induced in metallic inclusions. If the spatial scale of the system is smaller than the incident wavelength, the magnetic moments of individual current loops (within a single inclusion or formed by a number of them) give rise to magnetisation and the effective permeability.[2,22,24] The interaction between induced currents can be considered within the effective medium approximation. Composite materials with inclusions of a complex form (split rings, chiral and omega partials) are known to have unusual magnetic properties including both giant paramagnetic effect and negative $\mu_{eff}$.[10-12] These properties are related to resonance interaction of the electromagnetic wave with an inclusion and have been reported for microwave frequencies. The complex form of the inclusion is needed to realise the resonant conditions at wavelengths much larger then the inclusion size. For example, in a unit with two coaxial rings having oppositely oriented splits[12] a large capacitance is generated lowering the resonance frequency considerably. When the inclusion dimensions are reduced down to nanoscale, the wavelength can be proportionally decreased down to microns falling in the optical range. Then, the effective permeability of nanocomposites can be substantial at optical frequencies. However, it seems that the fabrication of nano-inclusions of a complex form may not be a realistic task. Fortunately, in the optical region the resonance frequency is lowered due to localised plasmon modes and fairly simple inclusions of a loop shape will create substantial magnetic moments even when the inclusion size is much smaller than the wavelength of the incident light.

Within the effective medium theory, the problem is reduced to considering the scattering of electromagnetic wave by a metallic contour. In general, this process can be very complicated. Here we consider two types of inclusions: ring-shaped inclusions (having some gap large enough to neglect the edge capacitance) and pairs of parallel conducting sticks connected via displacement currents (see Fig. 1), which permit a fairly simple analysis at certain approximations. For the chosen geometry, the resonance of localised plasmons confined within the inclusion is realised, which leads to the resonant current distributions and eventually, to large induced magnetic moments. The latter is responsible for the effective permeability having a resonance-like dispersion law with the resonant frequency coinciding with that for the current distribution.

Let us formulate the basic assumptions under which the problem is treated. We consider the composite medium irradiated by a plane-polarised electromagnetic wave of a single frequency $\omega$, so that the time dependence is of the form $\exp(-i\omega t)$. The contour length ($l$) is much larger than the cross-section size ($2a$). Then, when calculating fields in the surrounding space, the thickness of the contour can be neglected and the induced currents can be replaced by the effective linear currents $\mathbf{j}$. The current distribution inside the inclusion affects the scattered fields only via the boundary conditions imposed at the inclusion surface. The wavelength is also much larger than the cross section, $\lambda >> 2a$, but there is no restrictions on $l$ with respect to $\lambda$. The magnetic moments in the composite are induced when the incident light has the magnetic field directed perpendicular to the plane of a metallic contour as shown in Fig. 3. In the case of a circular contour (Fig. 3(a)), the field **H** induces a circumferential current $j_m(\theta)$ depending on the azimuthal angle $\theta$. Then, the magnetic moment **m** associated with this current is:

$$\mathbf{m} = \frac{1}{2c}\int(\mathbf{r}\times\mathbf{J}_m(\mathbf{r}))dV = \mathbf{o}_z \frac{R_0^{\,2}}{2c}\int_0^{\theta_0} j_m(\theta)d\theta\,, \tag{13}$$

where $\mathbf{J}_m = \mathbf{j}_m(\theta)\delta_S$, $\delta_S$ is the two-dimensional Dirac delta-function which peaks at the axis of the wire, $R_0$ is the radius of the contour, $(2\pi - \theta_0)$ is the angle of the gap, and $\mathbf{o}_z$ is the unit vector perpendicular to the contour plane. The currents due to the electric field give no contribution to **m**. In the case of a pair of conducting sticks (Fig. 3(b)) the current is distributed along their length and can form closed contours via the displacement currents. The metallic sticks as elements to produce the effective permeability were first proposed in Ref. 25. In this work, however, a random assembly of metallic sticks was considered, for which the total magnetic moment vanishes due to symmetry. As a result, the effective permeability for such system is unity, as was proven by experiments[26] (in Ref. 26 the response from a stick-composite is described adequately in terms of the effective permittivity, indicating that the effective permeability is essentially unity). The magnetic properties may appear only in diluted composites containing pairs of parallel sticks as a single element, as shown in Fig. 1(a). The magnetic moment is then found as (from symmetry, the contribution from the displacements currents equals to that from current $j_m(x)$):

$$\mathbf{m} = \mathbf{o}_z \frac{d}{c} \int_{-l/2}^{l/2} j_m(x) dx \,, \tag{14}$$

where $d$ is the distance between the sticks. The magnetic polarisability $\chi_0$ of a single inclusion associated with the induced moment can be found from $\mathbf{m} = \chi_0 V \mathbf{H}$ where $V$ is the volume of the metallic inclusion. The effective permeability is calculated from a self-consistent equation of the type: [2,22,24]

$$\mu_{eff} = 1 + 4\pi p \chi_0(\mu_{eff}) \,. \tag{15}$$

If the incident electromagnetic wave has the electric field **E** parallel to the wires (Fig. 1(a)) a substantial electric dipole moment is generated contributing to the effective permittivity. The currents $\mathbf{j}_e$ induced by this field can be considered separately as shown in the following Section. The electric dipole moment $\mathbf{p}$ and the dielectric polarisability $\alpha_0$ of the inclusion are calculated for stick contour using the continuity equation $\partial j_e / \partial x = i\omega \rho$ and integrating by parts with boundary conditions $j_e(\pm l/2) \equiv 0$ ( $\rho$ is the charge density per unit length):

$$\mathbf{p} = \mathbf{o}_x \frac{i}{\omega} \int_{-l/2}^{l/2} j_e(x) dx \,, \quad \mathbf{p} = \alpha_0 V \mathbf{E} \,. \tag{16}$$

The effective permittivity can be found from the self-consistent equation similar to (15).[25]

In our analysis it will be important to consider a resonance distribution of the induced currents. From the microwave antenna theory it is known that a nontrivial current distribution occurs when the wavelength is in the range of *l*. Then, the use of the effective dielectric and magnetic parameters is doubtful. However, in our case the situation is different. In optical and infrared spectral ranges, metal conductivity $\sigma$ can be approximated by Drude formula:

$$\sigma(\omega) = \frac{\sigma_0}{1 - i\omega\tau}, \tag{17}$$

where $\sigma_0 = \omega_p{}^2\tau / 4\pi$, $\omega_p$ is the plasma frequency, $\tau$ is the relaxation time (for silver, $\sigma_0 = 5.7\times10^{17}$ s$^{-1}$ and $\tau = 2.7\times10^{-14}$ s). In the high frequency range considered here ($\omega \sim 10^{15}$ s$^{-1}$) losses in metal grains are relatively small, $\omega\tau >> 1$. Therefore, the metal conductivity is characterised by the dominant imaginary part. This is very important for our analysis since the current can have a resonance for a considerably larger wavelength $\lambda >> 2l$. Physically, this is associated with the resonance of localised plasmon modes.

In (17) the relaxation time $\tau$ has a meaning of the mean-free time between electron collisions. The metal inclusions considered here have a length in the range of 100 nm and a cross-section size of 10 nm. The mean-free pass in noble metals such as silver is about 40 nm. It implies that the parameter $\tau$ used in (17) differs from in bulk materials. However, in the frequency range $\omega\tau >> 1$, electrons oscillate many times between collisions and the collisions are of little importance. The conductivity has a dominant imaginary part independent of $\tau$:

$$\sigma(\omega) = \frac{\omega_p^2}{4\pi\,\omega^2\tau} + i\frac{\omega_p^2}{4\pi\,\omega}$$

The resistive losses, which are determined by the real part, are typically smaller or in the range of the radiation losses and will only slightly change the values of currents at resonance. However, the resonance frequency will shift considerably towards higher frequencies if $\omega\tau \sim 1$. This imposes limitations on the minimum cross-section.

## V. Mathematical background

### A. Basic equations

Let us consider the current distribution in a thin metallic conductor irradiated by an electromagnetic field. The approximations used are: $2a << l$, $\lambda >> 2a$. This is a standard problem of the antenna theory (see, for example, Refs. 27, 28), which can be treated in terms of retarded scalar $\varphi$ and vector $\mathbf{A}$ potentials. The total electric field $\mathbf{E}_t = \mathbf{e}_0 + \mathbf{e}$ is represented by the sum of the external field $\mathbf{e}_0$ and the scattered field $\mathbf{e}$. In the Lorenz gauge $\varepsilon\,\partial\varphi/\partial t + 4\pi\ \mathrm{div}\,\mathbf{A} = 0$, the equation for $\mathbf{e}$ is written as:

$$\mathbf{e} = \frac{4\pi\, i\omega\mu}{c^2}\mathbf{A} - \frac{4\pi}{i\omega\varepsilon}\,\mathbf{grad}\,\mathbf{div}\mathbf{A}\,. \tag{18}$$

The vector potential $\mathbf{A}$ taken at arbitrary point $\mathbf{r}_0$ is obtained in the form of a convolution with the total current density $\mathbf{J}(\mathbf{r})$:

$$\mathbf{A}(\mathbf{r}_0) = (G * \mathbf{J}) = \int_V \mathbf{J}(\mathbf{r}')G(r)\,dV_{\mathbf{r}'}, \qquad G(r) = \frac{\exp(i\,kr)}{4\pi\, r}, \tag{19}$$

where $\mathbf{r} = \mathbf{r}_0 - \mathbf{r}'$, $r = |\mathbf{r}|$, integration is taken over the volume containing current, $k = (\omega/c)\sqrt{\varepsilon\mu}$ is the wave number, $G(r)$ is the Green function satisfying the Helmholtz equation. Customarily, equation (19) is solved under zero boundary condition:

$$\bar{\mathbf{E}}_t \equiv 0\,, \tag{20}$$

where $\bar{\mathbf{E}}_t$ is the tangential component of the total electric field taken at the surface of the conductor. Then, the current distribution is found from an integro-differential equation. The condition (20) corresponds to the case of an ideal conductor with infinite conductivity. Being used as an approximation, (20) works reasonably well when the radiation losses are considerably larger than the resistive ones or the system is out of resonance. However, in certain cases (including ours), the current distribution may have a zero (or greatly reduced) dipole moment. This implies that the radiation losses are comparable with the resistive ones and the condition (20) is no longer valid. The processes related to a finite conductivity may change the resonance condition for the current distribution: reduce the current amplitude and shift of the resonance wavelength. Here the problem is solved imposing impedance boundary conditions, which is valid at any frequency including the optical range:

$$\overline{\mathbf{E}}_t = \hat{\varsigma}(\overline{\mathbf{H}}_t \times \mathbf{n}), \tag{21}$$

where $\hat{\varsigma}$ is the surface impedance matrix, **n** is the unit vector normal to the surface and directed inside the conductor, $\overline{\mathbf{H}}_t$ is the tangential component of the total magnetic field taken at the surface (hereafter, overbar is used to denote tangential fields at the inclusion surface). For the geometry considered here (see Figs. 1,3), the external magnetic field is normal to the contour plane and gives no contribution to (21). The scattered field **h** is determined as:

$$\mathbf{h} = \frac{4\pi}{c}\,\mathbf{curl}\,\mathbf{A},$$

or,

$$\mathbf{h}(\mathbf{r}_0) = \frac{1}{c}\int_V \frac{(1 - i\,k\,r)\exp(i\,k\,r)}{r^3}(\mathbf{J}(\mathbf{r}')\times\mathbf{r})\,dV_{\mathbf{r}'}. \tag{22}$$

When the skin effect is strong, equation (21) does not depend upon geometry and for a nonmagnetic conductor (permeability of metal is always unity at optical frequencies) is represented by a scalar (normal skin-effect):

$$\varsigma = (1 - i)\sqrt{\frac{\omega}{8\pi\,\sigma}}. \tag{23}$$

The case of a thin arbitrarily shaped conductor having a circular cross section allows the surface impedance to be determined for any frequencies. The electromagnetic field inside such a conductor can be taken to be the same as that inside a straight cylinder. Then, in the local cylindrical coordinate system ($r$, $\varphi$, $x$) with the axis $x$ in the axial direction the impedance boundary conditions (21) become:

$$\bar{E}_{tx} = \varsigma_{xx} \bar{h}_{\varphi}$$
$$\bar{E}_{t\varphi} = -\varsigma_{\varphi\varphi} \bar{h}_{x}, \tag{24}$$

with[29]

$$\varsigma_{xx} = \frac{k_0 c}{4\pi\sigma} \frac{J_0(k_0 a)}{J_1(k_0 a)}, \qquad \varsigma_{\varphi\varphi} = -\frac{k_0 c}{4\pi\sigma} \frac{J_1(k_0 a)}{J_0(k_0 a)}, \tag{25}$$

where $k_0^2 = 4\pi i \sigma \omega / c^2$ and $J_0$, $J_1$ are the Bessel functions of the zero and first order, respectively. Equations (25) are valid for normal skin-effect. For the dimensions considered, the skin depth is in the range of the mean-free pass, and both these parameters are larger than $a$. Then, the skin effect is week, and using (25) is still reasonable.

Since we are interested only in fields in the surrounding space, the current inside a thin conductor can be replaced by an effective linear one $j(x)$ that flows along the axis of the wire: $\mathbf{J}(\mathbf{r}) = \mathbf{j}(x)\delta_S$. The volume integration in (19) is then replaced by the integration along the current contour. Thus, the current distribution in a thin conductor irradiated by the electromagnetic field is found from equations (18), (19) with boundary condition (24) that binds scattered fields **e** (18) and **h** (22) taken on the surface of the conductor.

## B. Current equation in a straight wire with circular cross section

First, the current equation is obtained for a straight wire with a circular cross section placed in the electrical field $\mathbf{e}_0$ (of any origin) parallel to the wire axis $x$. The scattered field **e** is determined by the $x$-component of the vector potential. The value of the longitudinal electric field $\bar{e}_x(x)$ taken at the wire surface is represented in terms of the integro-differential operator with respect to $x$, as it follows from (18), (19):

$$\bar{e}_x(x) = -\frac{4\pi}{i\omega\varepsilon} \left[ \frac{\partial^2}{\partial x^2} (G * j) + k^2 (G * j) \right], \tag{26}$$

$$(G*j)=\int_{-l/2}^{l/2} j(s)G(r)ds\,, \qquad r=\sqrt{(x-s)^2+a^2}\,.$$

For this geometry, the scattered magnetic field $\bar{h}_\varphi$ taken at the surface is circumferential. In (22), considering that the effective linear current $j(x')$ is flowing along the wire axis and $\mathbf{r}_0$ points at the wire surface yields $(\mathbf{J}(x')\times\mathbf{r})|_\varphi = j(x')a$. Then, the equation for $\bar{h}_\varphi$ obtains the form:

$$\bar{h}_\varphi(x)=\frac{2}{ac}(G_\varphi * j)=\frac{2}{ac}\int_{-l/2}^{l/2} j(s)G_\varphi(r)ds\,, \tag{27}$$

where $G_\varphi(r)=a^2(1-ikr)\exp(ikr)\big/2r^3$.

Finally, substituting (26) and (27) into boundary condition (24) yields the integro-differential equation for the linear current $j(x)$:

$$\frac{\partial^2}{\partial x^2}(G*j)+k^2(G*j)=\frac{i\omega\varepsilon}{4\pi}\bar{e}_{0x}(x)-\frac{i\omega\varepsilon\,\varsigma_{xx}}{2\pi ac}(G_\varphi * j)\,. \tag{28}$$

Equation (28) is solved imposing zero boundary conditions at the wire ends: $j(-l/2)=j(l/2)=0$ (the end surfaces are assumed small and associated capacitance is neglected).

Real parts of the Green functions $\mathrm{Re}(G)$ and $\mathrm{Re}(G_\varphi)$ have sharp peaks at $s=x$, which makes it possible to use the following approximations for calculating the convolutions:[30]

$$(\mathrm{Re}(G)*j)\approx j(x)\int_{-l/2}^{l/2}\mathrm{Re}(G(r))ds = j(x)Q\,,$$

$$Q=\int_{-l/2}^{l/2}\mathrm{Re}(G(r))ds \propto \frac{1}{4\pi}\int_{-l/2}^{l/2}\frac{ds}{\sqrt{s^2+a^2}}\sim\frac{\ln(l/a)}{2\pi}\,, \tag{29}$$

$$(\mathrm{Re}(G_\varphi)*j) \approx j(x)\int_{-l/2}^{l/2}\mathrm{Re}(G_\varphi(r))ds = j(x)Q_\varphi \,,$$

$$Q_\varphi = \int_{-l/2}^{l/2}\mathrm{Re}(G_\varphi(r))ds \propto \frac{a^2}{2}\int_{-l/2}^{l/2}\frac{ds}{(s^2+a^2)^{3/2}} + \frac{a^2k^2}{2}\int_{-l/2}^{l/2}\frac{ds}{\sqrt{s^2+a^2}} \propto (1+a^2k^2\ln(l/a)) \sim 1$$

where $Q$ and $Q_\varphi$ are positive form-factors. The logarithmic term in $Q_\varphi$ is neglected since $ak \ll 1$ in our case.

The convolutions $j(x)$ with $\mathrm{Re}(G)$ and $\mathrm{Re}(G_\varphi)$ give the main contribution to equation (28): $|(\mathrm{Im}(G)*j)| \ll |(\mathrm{Re}(G)*j)|$ and $|(\mathrm{Im}(G_\varphi)*j)| \ll |(\mathrm{Re}(G_\varphi)*j)|$. On the other hand, convolutions $j(x)$ with imaginary parts $\mathrm{Im}(G)$ and $\mathrm{Im}(G_\varphi)$ are responsible for radiation losses and become important at resonance. They can be calculated by the iteration method given in Appendix A. Equation (28) is reduced to an ordinary differential equation for the zero approximation $j_0(x)$ where the radiation losses are neglected:

$$\frac{\partial^2}{\partial x^2} j_0(x) + k_1^2 j_0(x) = \frac{i\omega\varepsilon}{4\pi Q}\bar{e}_{0x}(x), \tag{30}$$

$$k_1 = kg\,, \tag{31}$$

$$g = \left(1 + \frac{ic\varsigma_{xx}}{2\pi a\omega\mu}\frac{Q_\varphi}{Q}\right)^{1/2} \sim \left(1 + \frac{ic\varsigma_{xx}}{a\omega\mu\ln(l/a)}\right)^{1/2}.$$

Equation (31) shows that the impedance boundary condition renormalizes the wave number of the incident radiation. Considering the solution of (30) with $j(-l/2) = j(l/2) = 0$, the resonance wavelengths are determined via this new wave number from the condition $\cos k_1 l/2 = 0$ or $k_1 l = \pi(2n-1)$ :[27,28]

$$\lambda_{res,n} = \frac{2l}{2n-1}\mathrm{Re}(g\sqrt{\varepsilon\mu}), \quad n = 1,\ 2,\ 3... \tag{32}$$

A similar renormalisation method for the wave number has been used to tackle boundary effects in the microwave scattering from a conducting stick placed in a thin dielectric layer.[31]

### C. Current equation for two parallel wires

With the help of equation (28), we can now consider the current distribution in two parallel wires. The distance $d$ between them has to be larger than the diameter ($d > 2a$) in order to use the approximation of thin conductors. The equations for currents are of the form:

$$\frac{\partial^2}{\partial x^2}(G * j_1) + k^2 (G * j_1) = \frac{i\omega\varepsilon}{4\pi}(\bar{e}_{01x}(x) + \bar{e}_{21x}(x)) - \frac{i\omega\varepsilon\varsigma_{xx}}{2\pi a c}(G_\varphi * j_1), \quad (33)$$

$$\frac{\partial^2}{\partial x^2}(G * j_2) + k^2 (G * j_2) = \frac{i\omega\varepsilon}{4\pi}(\bar{e}_{02x}(x) + \bar{e}_{12x}(x)) - \frac{i\omega\varepsilon\varsigma_{xx}}{2\pi a c}(G_\varphi * j_2).$$

Here $\bar{e}_{12x}$ and $\bar{e}_{21x}$ are the longitudinal electric fields induced by each conductor at the surface of other, $\bar{e}_{01x}$ and $\bar{e}_{02x}$ are the external fields, $j_1(x)$ and $j_2(x)$ are the linear currents inside the conductors. The fields $\bar{e}_{12x}$ and $\bar{e}_{21x}$ are determined from (26) with $r = r_d = \sqrt{(x-s)^2 + d^2}$ .

Equations (33) are reduced to two independent equations by introducing $j_e = (j_1 + j_2)/2$ and $j_m = (j_1 - j_2)/2$:

$$\frac{\partial^2}{\partial x^2}((G - G_d) * j_m) + k^2((G - G_d) * j_m) = \frac{i\omega\varepsilon}{8\pi}(\bar{e}_{01x} - \bar{e}_{02x}) - \frac{i\omega\varepsilon\varsigma_{xx}}{2\pi a c}(G_\varphi * j_m), \quad (34)$$

$$\frac{\partial^2}{\partial x^2}((G + G_d) * j_e) + k^2((G + G_d) * j_e) = \frac{i\omega\varepsilon}{8\pi}(\bar{e}_{01x} + \bar{e}_{02x}) - \frac{i\omega\varepsilon\varsigma_{xx}}{2\pi a c}(G_\varphi * j_e), \quad (35)$$

$G_d(r_d) = \exp(ikr_d)/4\pi r_d$ .

It is easy to see that only electric field $e_d$ directed along the wires and magnetic field $H$ perpendicular to the two-wire contour (*xy*-plane) will excite currents inside the inclusion in the discussed geometry. The perpendicular magnetic field produces circulatory electric field $e_m = (i\omega/2c)\mu H l d/(l+d)$ along the wire contour due to Faraday's law of induction. Therefore, the external electric fields are of the form: $\bar{e}_{01x} = e_d + e_m$ and $\bar{e}_{02x} = e_d - e_m$ . This implies that the current $j_m$ in the first equation is induced by the magnetic field and is responsible for the magnetic moment $\mathbf{m}$. The current $j_e$ entering the second equation is created by the electric field and defines the electric dipole moment $\mathbf{p}$.

For $d$ sufficiently small (strong interaction), the convolution with the Green function $G_d(r_d)$ can be estimated with the help of an approximate formula (29):

$$(\mathrm{Re}(G_d)*j) \approx j(x)Q_d, \tag{36}$$

$$Q_d = \int_{-l/2}^{l/2} \mathrm{Re}(G_d(r_d))ds \propto \frac{1}{4\pi}\int_{-l/2}^{l/2} \frac{ds}{\sqrt{s^2+d^2}} \sim \frac{\ln(l/d)}{2\pi}.$$

Similar to equation (30), the zero approximation for the current distribution reads:

$$\frac{\partial^2}{\partial x^2} j_{m0}(x) + k_m^2 j_{m0}(x) = \frac{i\omega\varepsilon}{4\pi(Q-Q_d)} e_m, \tag{37}$$

$$\frac{\partial^2}{\partial x^2} j_{e0}(x) + k_e^2 j_{e0}(x) = \frac{i\omega\varepsilon}{4\pi(Q+Q_d)} e_d, \tag{38}$$

$$k_m = kg_m,\ g_m = \left(1 + \frac{ic\varsigma_{xx}}{2\pi a\omega\mu}\frac{Q_\varphi}{(Q-Q_d)}\right)^{1/2} \sim \left(1 + \frac{ic\varsigma_{xx}}{a\omega\mu\ln(d/a)}\right)^{1/2}, \tag{39}$$

$$k_e = kg_e,\ g_e = \left(1 + \frac{ic\varsigma_{xx}}{2\pi a\omega\mu}\frac{Q_\varphi}{(Q+Q_d)}\right)^{1/2} \sim \left(1 + \frac{ic\varsigma_{xx}}{a\omega\mu\ln(l^2/da)}\right)^{1/2}. \tag{40}$$

The general solution of (37), (38) is of the form:

$$j_{m,e0}(x) = A\sin(k_{m,e}x) + B\cos(k_{m,e}x) + \frac{i\omega\varepsilon e_{m,d}}{4\pi(Q\mp Q_d)k_{m,e}^2}. \tag{41}$$

In (41), the first set of subscripts corresponds to the magnetic excitation for which sign "minus" in the last term is taken, and the second set corresponds to the electric excitation with sign "plus". Imposing zero boundary condition ($j_{m,e0}(-l/2) = j_{m,e0}(l/2) \equiv 0$) yields:

$$j_{m,e0}(x) = \frac{i\omega\varepsilon e_{m,d}}{4\pi(Q\mp Q_d)k_{m,e}^2}\frac{(\cos(k_{m,e}l/2) - \cos(k_{m,e}x))}{\cos(k_{m,e}l/2)}. \tag{42}$$

The resonance wavelengths $\lambda_m$, $\lambda_e$ for the two excitations are different and can be found from the condition $k_{m,e}\, l = \pi(2n-1)$:[27,28]

$$\lambda_{m,e} = \frac{2l}{2n-1}\mathrm{Re}(\sqrt{\varepsilon\mu}\, g_{m,e}), \quad n = 1,\ 2,\ 3... \tag{43}$$

The amplitude of the current at resonance is restricted by losses related to conductivity and relaxation properties of the surrounding medium. However, solution (41) does not contain such an important factor as radiation losses since the imaginary parts of convolutions were neglected. To find the effect of radiation, original equations (33) can be solved by iterations. For this purpose, it is converted to an integral form in the Appendix A. The *n*-th iteration can be represented as:

$$j_n(x) = j_0(x) + \int_{-l/2}^{l/2} S(x,q) j_{n-1}(q) dq\,, \tag{44}$$

where the zero-iteration $j_0$ has to be taken in the form of (41) since the boundary condition is imposed for $j_n$ and $S(x,q)$ is the integral kernel. The iteration method is proved to converge very rapidly. The first iteration $j_1$ is sufficient to take account of the radiation effects. Its explicit form and the form of the kernel $S(x,q)$ are calculated in Appendix A.

The scattering at a ring inclusion gives similar results for the current distribution and magnetic polarisability. This case is considered in Appendix B.

## VI. Results and discussion

We are now in a position to proceed with the analysis of the effective magnetic and electric properties associated with the currents induced in the metallic contours. The characteristic size *l* is taken to be in the range of 100 nm and the cross-sectional size of 10 nm. These scales can be achieved in practice. The conductivity obeying the Drude equation (17) with parameters typical of such noble metals as silver and copper are used in all the calculations. The effect of a smaller relaxation time (in comparison with that of a bulk metal) due to electron scattering at the inclusion surface results in shifting the dispersion region to higher frequencies.

**Two-wire contour**

First, we consider two-wire contour composites irradiated by light having the magnetic field $H$ perpendicular to the contour plane and the electric field $E$ perpendicular to the wires. For this polarisation, the circulatory currents $j_m$ are induced leading to the effective permeability. The effective permittivity can be considered to be unity. The dimensions used are as follows: $l = 100$ nm, $a = 6$ nm, $d = 30$ nm. It is useful to investigate the forms of the current distribution for different frequencies along with the dispersion law of the magnetic polarisability of the inclusion $\chi_0 = \chi' + i\chi''$. Figure 4 shows plots of $j(x) = j'(x) + i\, j''(x)$ as a function of a distance $x$ along the wire for three frequencies: $f < f_{res}$, $f \sim f_{res}$, and $f > f_{res}$ where $f_{res} = c / \lambda_{res}$ is the resonance frequency. The current distribution is calculated using formulae (37) and (A7) obtained in the zero approximation and using first iteration, respectively. In this case, the resonance wavelength is found to be $\lambda_{res} = 730$ nm, which is several times larger than it could be expected from half-wavelength resonance condition ($\lambda_{res} = 2l$) known for microwave antennas.[27,28] The skin effect is weak for the chosen wire radius and the contribution from the surface impedance causes this remarkable shift of the resonance wavelength since the conductivity is dominantly imaginary for these frequencies. Physically, this result corresponds to localised plasmon modes inside the wire.

On the other hand, the current distribution exhibits all the features typical of those in microwave antennas. The real part $j'$ changes phase at $f_{res}$. For frequencies below the resonance (Fig. 4(a)) $j'$ is positive and magnetic polarisability of the inclusion $\chi_0$ exhibits a paramagnetic response, as it is seen in Fig. 5. For $f > f_{res}$ (Fig. 4(b)), $j'$ is negative and $\chi_0$ is of a diamagnetic character, showing quite large negative values. Closer to the resonance, $j'$ undergoes rapid transformations (Fig. 4(c)). In this frequency range small factors such as radiation may introduce essential changes in $j'(x)$ plots, but the integral parameter $\chi_0(\omega)$ does not change much showing only a small shift of the resonance frequency and a slight decrease in the resonance peaks. For the considered geometry, the dielectric dipole moment is zero and the radiation is strongly reduced.

Figure 6 shows the effective permeability $\mu_{eff} = \mu' + i\mu''$ calculated for the volume concentration of inclusions $p = 6\%$ by considering the current contours as independent magnetic moments (dipole sum) and within the effective medium theory (EMT) (see Eq. (15)). In this case, the wire radius was chosen to be 10 nm to demonstrate that the resonance shifts to higher frequencies since the skin effect is stronger for a larger cross-section. For non-interacting moments, the resonance peaks in permeability are large reaching negative values past the resonance and the dispersion region is narrow, whereas the interaction broadens the permeability behaviour and reduces the peaks of $\mu'$ and $\mu''$ greatly, so that the real part is always positive. An important characteristic is that $\mu'$ peaks at lower frequency when $\mu''$ is very small. Thus, there exists a range of frequencies where the light propagation is affected by magnetic properties but the light absorption is still negligible. It may happen that EMT is a rather rough approximation for the considered system. A periodic array of two-wire contours may exhibit a stronger magnetic activity as suggested by the dipole sum result, which do show negative $\mu'$. In any way, the actual behaviour is somewhere in between the considered cases and the least range of $\mu'$ variation is 1.5–0.5 for this concentration, which is quite big. It has to be noted that the magnetic properties in two-wire contour system cannot be enhanced by simply increasing the concentration. If the distance between the pairs is in the range of $d$, then the induced magnetic properties disappear as it is clear from symmetry. The system of randomly placed wires has no magnetic properties (neglecting those due to circumferential currents inside the wires), as already discussed in Section IV.

We now consider the polarisation of the incident light for which $H$ is still perpendicular to the contour plane but $E$ is along the wires. For this case, both $\mu_{eff}$ and $\varepsilon_{eff}$ are essential. The result for $\mu_{eff}$ is the same as considered previously since the currents $j_m$ and $j_e$ due to $H$ and $E$ contribute independently to the magnetic and electric polarisabilities, respectively. The current $j_e$, which determines the electric polarisability of the inclusion $\alpha_0$, is calculated in zero approximation (38) and taking first iteration (A7). In this case, the results for $\alpha_0 = \alpha' + i\alpha''$ differ greatly for the zero and first approximations, as shown in Fig. 7, because of a substantial radiation effect. The polarisability has resonant dispersion behaviour. The radiation losses make the resonance wider, shift the resonance frequency and reduce the resonant peaks. Comparing plots $\alpha_0(\omega)$ and $\chi_0(\omega)$ (Fig. 5) it is seen that the resonance frequency for $\alpha_0$ is higher, since the resonance for $j_e$ happens at higher frequency than for $j_m$. The effective permittivity $\varepsilon_{eff} = \varepsilon' + i\varepsilon''$ calculated for $p = 3\%$ within EMT (see Eq. (15)) shows a very broad dispersion region as seen in Fig. 8. In terms of $\varepsilon_{eff}$, the effect of radiation is not pronounced since the interaction itself has a similar effect of smoothing the resonance characteristics. The real part of the permittivity is negative near the high-frequency side of the resonance. Figure 9 compares the dispersion behaviour for $\varepsilon_{eff}$ and $\mu_{eff}$ ($p = 3\%$) for the aforementioned polarization. The resonance region for $\varepsilon_{eff}$ is shifted towards higher frequencies. In the area of magnetic resonance, $\mu_{eff} \sim \varepsilon_{eff}$ with very small losses: $\varepsilon'' << 1, \mu'' << 1$. It means that inequality (12) is satisfied since $l/\lambda << 1$ and the concept of permeability is meaningful at optical frequencies. Therefore, this system can be useful for designing materials with effective parameters suitable for new optical effects described in Section II.

**B. Ring-contour**

A similar magnetic behaviour is obtained for ring-composite materials irradiated by light having the magnetic field perpendicular to the plane of the ring (see Appendix B). The electric field of the incident light is in the plane of the ring and always induces some effective permittivity. In this case, the magnetic and electric resonance frequencies coincide. The value of $\varepsilon_{eff}$ is reduced in comparison with that of the two-wire case since the average exciting electric field for the electric dipole moment is smaller. Then, $\mu_{eff} \sim \varepsilon_{eff}$ at resonance in this system.

Figure 10 shows the dispersion behaviour of the magnetic polarisability of the ring inclusion. The calculations are made with $R_0 = 50$ nm, $a = 5$ nm, and $\theta_0 = 320°$. The gap in the ring is sufficiently large to avoid any effects from the edge capacitance, which could be difficult to control at nano-scales. For these dimensions, the resonance wavelength is about 2 μm (infrared part of the spectrum). The radiation losses are essential because of the existence of the electric dipole moment. They strongly reduce the resonance peaks. In order to move the dispersion region to the visible spectral range, the ring diameter has to be decreased. However, we do not have much flexibility since for our analysis the condition $R_0 >> a$ is important. Further decrease in $a$ may be not realistic and will bring about complex behaviour of the conductivity in low dimensional systems. Taken $R_0 = 30$ nm, the resonance wavelength decreases down to 1.30 μm, which is still in the infrared. Surprisingly, the magnetic polarisability of a ring for these higher frequencies is substantially decreased due to strong radiation losses, which will result in much smaller values of the effective permeability. It appears that the dispersion region of $\mu_{eff}$ in ring-composites is essentially limited by infrared spectral range. The effective permeability for $R_0 = 30$ nm and two concentrations $p = 6\%$ and $p = 30\%$ is presented in Fig. 11. Similar to the two-wire composite, at frequencies where the real part has peaks, the imaginary part is small, which is important for possible applications. For ring-composites, the concentration can be increased, which allows the negative permeability to be realised, as demonstrated in Fig. 11(b).

**Conclusion.**

We have shown that a metal-dielectric composite having loop-shape nanoscale inclusions responds to optical radiation as if it has effective magnetic properties. Such material can be named as optomagnetic. It is known that the macroscopic magnetic properties originated by localised electrons in atom have no physical meaning from optical frequencies onward. In contrast, the effective permeability of the proposed composite is proven to be consistent with the macroscopic Maxwell equations even at optical frequencies, having values that can differ substantially from unity within a dispersion band. New optical effects are predicted, which are related with specific conditions of reflection/refraction at interface with such a medium. They are likely to find applications in optical filters, sensors, polarizes and other optoelectronics devices. An interesting example is the reflectionless normal incidence from vacuum when the permeability and permittivity are the same. This condition is an optical analogy of the impedance matching known for microwaves and is quite realistic in the considered optomagnetic materials for a certain narrow frequency range. In addition, the losses (imaginary parts of these parameters) can be small at those frequencies.

The analytical approach developed is based on solving the scattering problem for metallic inclusions of two types: a ring with a relatively large gap and a pair of parallel wires.

The method allows us to find the current and charge distributions within the inclusion, which constitute the effective permeability and permittivity. The localised plasmon modes are proven to play an important role as they make the microstructure to be resonant at frequencies much lower than those following from the half-wavelength requirement for microwave antennas. For example, the effective permeability of composites having two-wire inclusions of 100 nm long shows resonance behaviour with a characteristic frequency of $4 \cdot 10^{14}$ Hz (750 nm). The parameters determining the optical conductivity such as relaxation time (mean-free time between collisions) are also important to realise favourable resonance conditions. The use of noble metals as Au, Ag, Cu is preferable to increase the relaxation time. However, because the composite structure has nano-dimensions, the bulk parameters may need to be modified. This factor requires further investigation.

For composite with volume fraction more than 1% the interactions between inclusions become important. They are considered in a self-consistent manner using the effective medium theory. It turns out that the interactions broaden the dispersion region and strongly reduce the permeability peaks near the resonance, preventing it from having negative values. It may be that the effective medium theory for the considered system is a rough approximation. Then, the analysis of the effective permeability in a periodic array of loop-shaped inclusions allowing an exact solution would be of a considerable interest.

The analysis predicts that inherent metallic microstructure properties will limit magnetic activity of the type considered here by visible spectral range. More specifically, magnetic properties of the composites containing ring-shape inclusions will not be essential past the infrared as radiation effects become very strong. The radiation factor is reduced for a contour formed by two parallel wires. With this structural element the effective magnetism can exist in visible spectral range.

Finally, the resonant properties of the proposed optomagnetic medium strongly depend on conductivity. It is known that the conductivity of nanoinclusions can be changed considerably by external parameters such as bias magnetic or electric fields.[34,35] This opens up a possibility to create adaptive optics: modulators, tuneable lenses, and filters having small energy losses.

## IV. Appendix A.

Here we describe the iteration method of solving equations (34), (35). The convolutions of the current $j_{m,e}(x)$ with the Green functions are considered separately for the real and imaginary parts. Approximates (29) and (36) are used for the real parts. Equations (34), (35) with $j(x) = j_{m,e}(x)$ and $e_0(x) = \overline{e}_{01x} \mp \overline{e}_{02x}$ can be rewritten in the form:

$$\left[\frac{\partial^2}{\partial x^2} + \tilde{k}^2\right]\left[j(x) + \frac{i}{(Q \mp Q_d)}(\mathrm{Im}(G \mp G_d) * j)\right] = \\ = \frac{i\omega\varepsilon}{8\pi(Q \mp Q_d)} e_0(x) + \frac{i(\tilde{k}^2 - k^2)}{(Q \mp Q_d)}(\mathrm{Im}(G \mp G_d) * j) + \frac{\omega\varepsilon\varsigma_{xx}}{2\pi\, a c(Q \mp Q_d)}(\mathrm{Im}(G_\varphi) * j), \tag{A1}$$

where $\tilde{k}^2 = k_{m,e}^2 = \dfrac{\omega^2}{c^2}\varepsilon\mu\left(1 + \dfrac{i c \varsigma_{xx} Q_\varphi}{2\pi\, a\omega\mu\,(Q \mp Q_d)}\right)$.

The members of equation (A1) are grouped in a way to separate a renormalized wave number $\tilde{k}$. Equation (A1) can be treated as an inhomogeneous differential equation with respect to $\partial^2/\partial x^2 + \tilde{k}^2$. The general solution of this equation is represented by:

$$j(x) = A\sin(\tilde{k}x) + B\cos(\tilde{k}x) + \frac{i\omega\varepsilon}{8\pi(Q \mp Q_d)\tilde{k}} \int_{-l/2}^{x} \sin(\tilde{k}(x-s)) e_0(s) ds + \\ + \frac{i(\tilde{k}^2 - k^2)}{(Q \mp Q_d)\tilde{k}} \int_{-l/2}^{x} \sin(\tilde{k}(x-s))(\mathrm{Im}(G \mp G_d) * j)) ds + \\ + \frac{\omega\varepsilon\varsigma_{xx}}{2\pi\, a c(Q \mp Q_d)\tilde{k}} \int_{-l/2}^{x} \sin(\tilde{k}(x-s))(\mathrm{Im}(G_\varphi) * j) ds - \frac{i}{(Q \mp Q_d)}(\mathrm{Im}(G \mp G_d) * j). \tag{A2}$$

The parameters $A$ and $B$ are found from the boundary condition:

$$j(-l/2) = j(l/2) \equiv 0\,. \tag{A3}$$

Equation (A2) is the Fredholm equation of the second kind, which allows the iteration method to be successfully used with a rapid convergence. The zero iteration is constructed as follows:

$$j_0(x) = A\sin(\tilde{k}x) + B\cos(\tilde{k}x) + \frac{i\omega\varepsilon}{8\pi(Q\mp Q_d)\tilde{k}}\int_{-l/2}^{x}\sin(\tilde{k}(x-s))e_0(s)ds\,. \tag{A4}$$

The $n$-th iteration is can be written as:

$$\begin{aligned} j_n(x) = j_0(x) &+ \frac{i(\tilde{k}^2-k^2)}{(Q\mp Q_d)\tilde{k}}\int_{-l/2}^{x}\sin(\tilde{k}(x-s))(\mathrm{Im}(G\mp G_d)*j_{n-1}))ds + \\ &+ \frac{\omega\varepsilon\varsigma_{xx}}{2\pi\,ac(Q\mp Q_d)\tilde{k}}\int_{-l/2}^{x}\sin(\tilde{k}(x-s))(\mathrm{Im}(G_\varphi)*j_{n-1})ds - \frac{i}{(Q\mp Q_d)}(\mathrm{Im}(G\mp G_d)*j_{n-1}). \end{aligned} \tag{A5}$$

The parameters $A$ and $B$ have to be calculated at the final stage of the iteration method. In the case of the zero approximation, equation (A4) together with (A3) yield expression (42) of the main text ($e_m = const$). When the next iteration is considered, the parameters $A$ and $B$ are needed to be calculated again to satisfy (A3).

Introducing the integral kernel $S$ for equation (A5) gives:

$$j_n(x) = j_0(x) + \int_{-l/2}^{l/2} S(x,q)j_{n-1}(q)dq\,. \tag{A6}$$

Here:

$$j_0(x) = A\sin(\tilde{k}x) + B\cos(\tilde{k}x) + \frac{i\omega\varepsilon e_0}{8\pi(Q\mp Q_d)\tilde{k}^2},$$

$$S(x,q) = S_1(x,q) + S_2(x,q) + S_3(x,q),$$

$$S_1(x,q) = -\frac{i}{(Q\mp Q_d)}\mathrm{Im}(G(r)\mp G_d(r_d)),\; r = \sqrt{(x-q)^2+a^2}\,,\; r_d = \sqrt{(x-q)^2+d^2}\,,$$

$$S_2(x,q) = \frac{i(\tilde{k}^2-k^2)}{(Q\mp Q_d)\;\tilde{k}}\int_{-l/2}^{x}\sin(\tilde{k}(x-s))\,\mathrm{Im}(G(r)\mp G_d(r_d))ds\,, \qquad r = \sqrt{(s-q)^2+a^2}\,,$$

$$r_d = \sqrt{(s-q)^2+d^2}\,,$$

$$S_3(x,q)=\frac{\omega\varepsilon\varsigma_{xx}}{2\pi\, a\, c(Q\mp Q_d)\tilde{k}}\int\limits_{-l/2}^{x}\sin(\tilde{k}(x-s))\,\mathrm{Im}(G_\varphi(r))ds\,,\; r=\sqrt{(s-q)^2+a^2}\;.$$

The kernel S is written as the sum of three terms that represent three different sources of radiation. $S_1$ is a local kernel depending only on the wave number $k$ in free space. This contribution corresponds to that of the wire with infinite conductivity. The next two terms are non-local. $S_2$ is responsible for the radiation into free space partly penetrating back to the wires. Points in space are electrically bound via the conductors, which is represented by the convolutions with function $\sin(\tilde{k}(x-s))$. $S_3$ accounts for retarding effects related to impedance boundary condition. All three members of kernel $S$ contain a small factor $1/4\pi(Q\mp Q_d)$ resulting in a rapid convergence of the iteration sequence.

The first iteration gives:

$$\begin{aligned}j_1(x)=A\left(\sin(\tilde{k}x)+\int\limits_{-l/2}^{l/2}S(x,q)\sin(\tilde{k}q)dq\right)+B\left(\cos(\tilde{k}x)+\int\limits_{-l/2}^{l/2}S(x,q)\cos(\tilde{k}q)dq\right)+\\+\frac{i\omega\varepsilon e_0}{8\pi(Q\mp Q_d)\tilde{k}^2}\left(1+\int\limits_{-l/2}^{l/2}S(x,q)dq\right)\end{aligned}\;. \quad \text{(A7)}$$

The parameters $A$ and $B$ are found by solving two linear equations ( $j_1(\pm l/2)\equiv 0$ ):

$$\begin{pmatrix}\sin(\tilde{k}l/2)+a_{11} & \cos(\tilde{k}l/2)+a_{12}\\ -\sin(\tilde{k}l/2)+a_{21} & \cos(\tilde{k}l/2)+a_{22}\end{pmatrix}\otimes\begin{pmatrix}A\\B\end{pmatrix}=\begin{pmatrix}C\\D\end{pmatrix}, \quad \text{(A8)}$$

where

$$\begin{pmatrix}a_{11} & a_{12}\\ a_{21} & a_{22}\end{pmatrix}=\begin{pmatrix}\int\limits_{-l/2}^{l/2}S(l/2,q)\sin(\tilde{k}q)dq & \int\limits_{-l/2}^{l/2}S(l/2,q)\cos(\tilde{k}q)dq\\ \int\limits_{-l/2}^{l/2}S_1(-l/2,q)\sin(\tilde{k}q)dq & \int\limits_{-l/2}^{l/2}S_1(-l/2,q)\cos(\tilde{k}q)dq\end{pmatrix}, \quad \text{(A9)}$$

$$\begin{pmatrix} C \\ D \end{pmatrix} = -\frac{i\omega\varepsilon e_0}{8\pi(Q \mp Q_d)\tilde{k}^2} \begin{pmatrix} 1 + \int\limits_{-l/2}^{l/2} S(l/2, q)dq \\ 1 + \int\limits_{-l/2}^{l/2} S_1(-l/2, q)dq \end{pmatrix}.$$

In (A9) the equality $S_{2,3}(-l/2, q) \equiv 0$ is used. (A9) is represented as:

$$\begin{pmatrix} a_{11} + a_{21} & 2\cos(\tilde{k}l/2) + a_{12} + a_{22} \\ 2\sin(\tilde{k}l/2) + a_{11} - a_{21} & a_{12} - a_{22} \end{pmatrix} \otimes \begin{pmatrix} A \\ B \end{pmatrix} = \begin{pmatrix} C + D \\ C - D \end{pmatrix}. \quad \text{(A10)}$$

From (A10), $A$ and $B$ are given by:

$$A = \frac{(C - D) + B(a_{22} - a_{12})}{2\sin(\tilde{k}l/2) + a_{11} - a_{21}} \quad \text{(A11)}$$

$$B = \left( \frac{C + D}{2} + \frac{(D - C)(a_{11} + a_{21})}{4\sin(\tilde{k}l/2) + 2(a_{11} - a_{21})} \right) \Bigg/ \left( \cos(\tilde{k}l/2) + \frac{(a_{22} - a_{12})(a_{11} + a_{21})}{4\sin(\tilde{k}l/2) + 2(a_{11} - a_{21})} + \frac{a_{12} + a_{22}}{2} \right)$$

The resonance wavelengths are calculated by putting to zero the real part of the denominator in the expression for *B*. The resonance peaks are determined by its imaginary part taken at the resonance wavelength.

## V. Appendix B

The current distribution in a thin conductor of arbitrary form was analysed by K. Mei.[32,33] The problem was solved using the Fredholm equation of the first kind, which does not contain explicitly the wave operator $\partial^2/\partial x^2 + k^2$. As a result, the iteration procedure cannot be applied to this equation and the overall analysis is very complicated. Fortunately, for a circular current loop the problem can be formulated using the methods developed here for a straight conductor.

We will use cylindrical coordinates $(\rho, \theta, z)$ with the origin in the centre of the loop as shown in Fig. B1. The loop has a small gap of a segmental angle. The dihedral angle $\theta$ is measured from the gap. The vector potential **A** taken at the point $P(\mathbf{r}_0)$ is represented as a contour integral along the current loop:

$$A(\mathbf{r}_0) = \int_L \mathbf{j}(\mathbf{r}_s) G(r)\, ds \qquad \text{(B1)}$$

where $\mathbf{r}_s = (R_0, \theta_s, 0)$ is the vector pointing to the current element, $r = |\mathbf{r}_0 - \mathbf{r}_s|$, $R_0$ is the radius of the current loop, and $\mathbf{j}(\mathbf{r}_s)$ is the linear current. Because of symmetry, the scattered fields are described by only one component $A_\theta$ of the vector potential:

$$A_\theta(\mathbf{r}_0) = \int_L j(s) G(r) \cos(\theta_{\mathbf{r}_0} - \theta_s) ds \qquad \text{(B2)}$$

where the integration is with respect to $s = R_0 \theta_s$ and $j(s) = |\mathbf{j}(s)|$. The electric field $e_\theta$ is equated as:

$$e_\theta(\mathbf{r}_0) = -\frac{4\pi}{i\omega\varepsilon}\left[\frac{1}{\rho^2}\frac{\partial^2}{\partial\theta^2} A_\theta + k^2 A_\theta\right]. \qquad \text{(B3)}$$

Equation (B3) taken at the loop ($\mathbf{r}_0 = (R_0, \theta, 0)$) becomes:

$$\overline{e}_\theta(v) = -\frac{4\pi}{i\omega\varepsilon}\left[\frac{\partial^2}{\partial v^2}(G * j) + k^2 (G * j)\right], \qquad \text{(B4)}$$

$$(G * j) = \int_0^l j(s) G(r) \cos(\theta - \theta_s)\, ds, \qquad \text{(B5)}$$

$$r = \sqrt{R_0^2 + (R_0 + a)^2 - 2R_0(R_0 + a)\cos(\theta - \theta_s)}\ .$$

Here $v = R_0\theta$, $l = R_0\theta_0$ is the length of non-closed loop ($\theta_0 < 2\pi$). Formally, equation (B4) is similar to equation (26) for $\overline{e}_x$ in a straight wire. To use the impedance boundary conditions (24) we have to find the circumferential magnetic field $\overline{h}_\varphi$ (in local cylindrical coordinates $(a, \varphi, v)$ with $v$ along the loop axis). A general form of the scattered magnetic field taken at the loop point $v$ is:

$$\mathbf{h}(v) = \frac{1}{c}\int_L \frac{(1-ik\,r)\exp(ik\,r)}{r^3}(\mathbf{j}(s)\times\mathbf{r})\,ds \tag{B6}$$

where $r = \sqrt{R_0^2 + (R_0+a)^2 - 2R_0(R_0+a)\cos(\theta-\theta_s)}$. The integration in (B6) is divided into two parts, one for which $r > \Delta$ and the other for which $r < \Delta$, where $\Delta$ is a distance small compared with $R_0$ but large compared with $a$. In the first integral, the contribution to $\overline{h}_\varphi(v)$ averaged over the wire circumference is estimated to be of the order $a/R_0^3$, which is small and can be neglected (see expression (29) for $Q_\varphi$). For the integration where $r < \Delta$, we can take $(\mathbf{j}\times\mathbf{r})_\varphi = j(s)a$ and $\overline{h}_\varphi(v)$ is equated similar to that for a straight wire:

$$\overline{h}_\varphi(v) = \frac{a}{c}\int_{v1}^{v2} \frac{(1-ik\,r)\exp(ik\,r)}{r^3}\,j(s)ds \tag{B7}$$

where $\Delta = |v2 - v1|$. Although the parameter $\Delta$ is chosen arbitrarily, we assume that the integration is bounded in the segment $R_0\psi$ with angle $\psi \approx 2\sqrt{2a/R_0}$, between the points $v1$ and $v2$ as shown in Fig. B2. Then, $\overline{h}_\varphi$ is expressed in terms of the convolution:

$$\overline{h}_\varphi(v) = \frac{2}{ac}(G_\varphi * j), \quad (G_\varphi * j) = \int_{v1}^{v2} G_\varphi(r)j(s)ds, \tag{B8}$$

where

$$G_\varphi(r) = \frac{a^2(1-ikr)\exp(ikr)}{2r^3}, \tag{B9}$$

$v1 = \max\left(0, (v - \sqrt{2aR_0})\right)$, $v2 = \min\left(\theta_0 R_0, (v + \sqrt{2aR_0})\right)$.

Now we are able to formulate the integro-differential equation for the current distribution in a circular loop. Substituting (B4) and (B8) into boundary condition (24) yields:

$$\frac{\partial^2}{\partial v^2}(G*j) + k^2(G*j) = \frac{i\omega\varepsilon}{4\pi}\overline{e}_{\theta 0}(v) - \frac{i\omega\varepsilon\,\varsigma_{vv}}{2\pi\,a\,c}(G_\varphi * j), \tag{B10}$$

$j(0) = j(l) \equiv 0$.

Here the convolutions are defined by (B5) and (B8). The external electric field $\overline{e}_{\theta 0}(v)$ is considered to be circular, which is induced by the external magnetic field perpendicular to the loop plane. As in the case of two-wire contour, the external electric field in the plane of the ring does not affect the magnetic moment. The axial component of the surface impedance $\varsigma_{vv} = \varsigma_{xx}$ is defined by (25). Equation (B10) formally is similar to (28), (33) and can be solved using the method developed. The convolutions are estimated as:

$$(\mathrm{Re}(G) * j) \approx j(v) \int_0^l \mathrm{Re}(G(r)) \cos(\theta - \theta_s) ds = j(v) Q,$$

$$Q = \int_0^l \mathrm{Re}(G(r)) \cos(\theta - \theta_s) ds \propto \frac{1}{4\pi} \int_0^l \frac{\cos(\theta_0 / 2 - \theta_s) ds}{\sqrt{(s - l/2)^2 + a^2}} < \frac{\ln(l/a)}{2\pi},$$

$$(\mathrm{Re}(G_\varphi) * j) \approx j(v) \int_{v1}^{v2} \mathrm{Re}(G_\varphi(r)) ds = j(v) Q_\varphi, \tag{B11}$$

$$Q_\varphi = \int_{v1}^{v2} \mathrm{Re}(G_\varphi(r)) ds \propto \frac{a^2}{2} \int_0^\Delta \frac{ds}{((s - \Delta/2)^2 + a^2)^{3/2}} +$$
$$+ \frac{a^2 k^2}{2} \int_0^\Delta \frac{ds}{\sqrt{(s - \Delta/2)^2 + a^2}} \propto (1 + a^2 k^2 \ln(\Delta / a)) \sim 1.$$

The iteration method is formulated as follows:

$$\begin{aligned} j_n(v) &= j_0(v) + \frac{i(\tilde{k}^2 - k^2)}{Q \tilde{k}} \int_0^v \sin(\tilde{k}(v - s))(\mathrm{Im}(G) * j_{n-1})) ds + \\ &+ \frac{\omega \varepsilon \varsigma_{vv}}{2\pi a c Q \tilde{k}} \int_0^v \sin(\tilde{k}(v - s))(\mathrm{Im}(G_\varphi) * j_{n-1}) ds - \frac{i}{Q} (\mathrm{Im}(G) * j_{n-1}) \end{aligned}, \tag{B12}$$

where

$$j_0(v) = A \sin(\tilde{k} v) + B \cos(\tilde{k} v) + \frac{i \omega \varepsilon \overline{e}_{\theta 0}}{4\pi Q \tilde{k}^2},$$

$$\tilde{k} = kg, \; g = \left( 1 + \frac{i c \varsigma_{vv}}{2\pi a \omega \mu} \frac{Q_\varphi}{Q} \right)^{1/2}.$$

Introducing a general kernel $S$ gives (B12) in the form:

$$j_n(v) = j_0(v) + \int_0^l S(v,q) j_{n-1}(q) dq, \quad \text{(B13)}$$

$$S(v,q) = S_1(v,q) + S_2(v,q) + S_3(v,q), \quad \text{(B14)}$$

$$S_1(v,q) = -\frac{i}{Q} \mathrm{Im}(G(r)) \cos(\theta - \theta_q),$$

$$r = \sqrt{R_0^2 + (R_0 + a)^2 - 2R_0(R_0 + a)\cos(\theta - \theta_q)},$$

$$S_2(v,q) = \frac{i(\tilde{k}^2 - k^2)}{Q\ \tilde{k}} \int_0^v \sin(\tilde{k}(v-s)) \cos(\theta_s - \theta_q) \mathrm{Im}(G(r)) ds,$$

$$r = \sqrt{R_0^2 + (R_0 + a)^2 - 2R_0(R_0 + a)\cos(\theta_s - \theta_q)},$$

$$S_3(v,q) = \frac{\omega \varepsilon \varsigma_{vv}}{2\pi\, a c Q \tilde{k}} \int_0^v \sin(\tilde{k}(v-s)) \mathrm{Im}(G_\varphi(r)) \Theta(s,q)\, ds,$$

$$r = \sqrt{R_0^2 + (R_0 + a)^2 - 2R_0(R_0 + a)\cos(\theta_s - \theta_q)}.$$

Here $\Theta(s,q) = \begin{cases} 1, & q \in [s1(s), s2(s)] \\ 0, & q \notin [s1(s), s2(s)] \end{cases}$ is the "cutting" function. This function takes into account that the integration with respect to $q$ is made in the interval $[s1, s2]$, where $s1 = \max\left(0, (s - \sqrt{2aR_0})\right)$ and $s2 = \min\left(\theta_0 R_0, (s + \sqrt{2aR_0})\right)$.

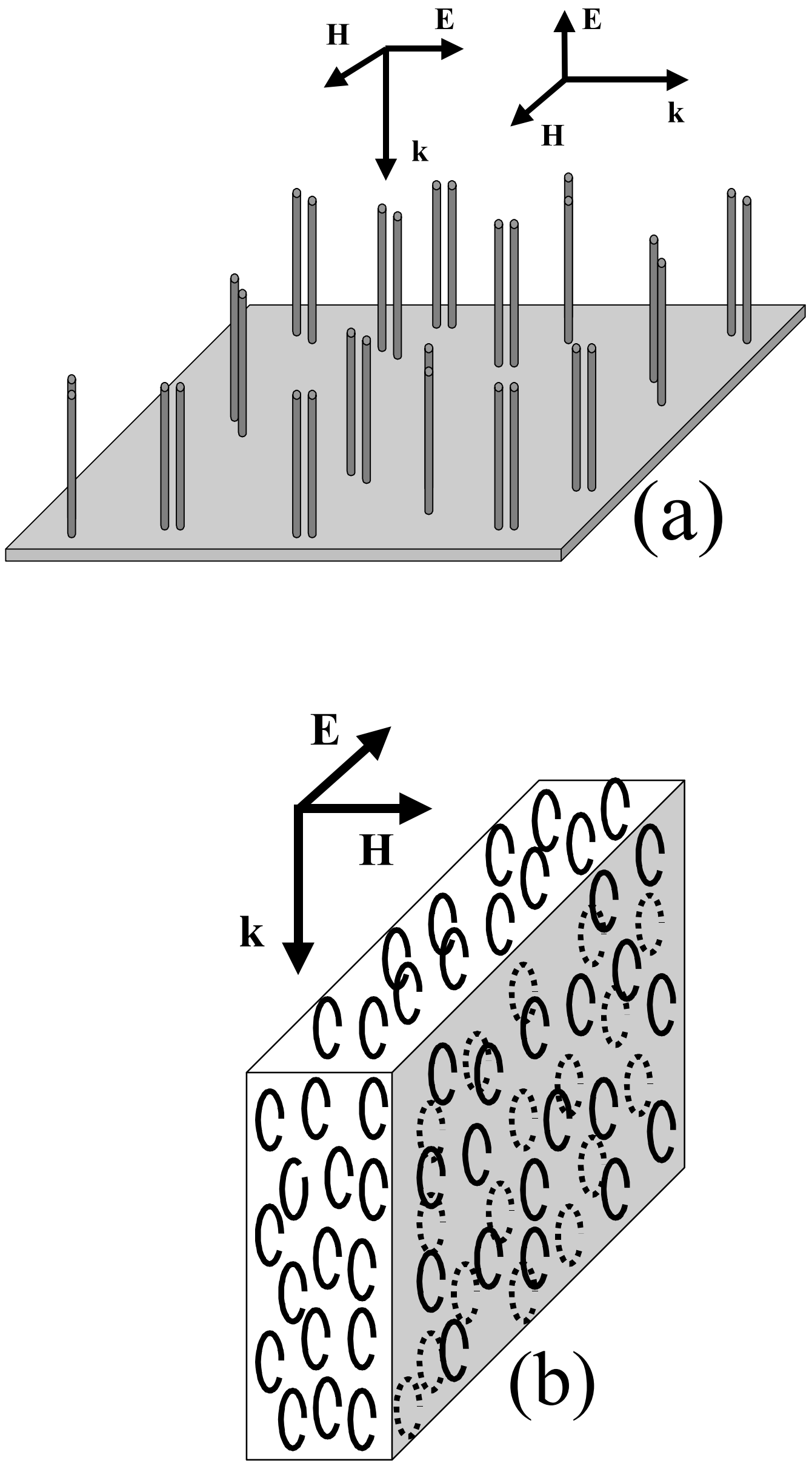

Fig. 1. Composite medium with effective permeability at optical frequencies. A possible polarisation of the electromagnetic wave associated with induced magnetic properties is indicated. In (a), a system of nanowires grown on a substrate is shown, in which two parallel wires form a constituent element. The spacing in the pair of wires is much smaller than the distance between the pairs. In (b), a nanoring-composite is shown.

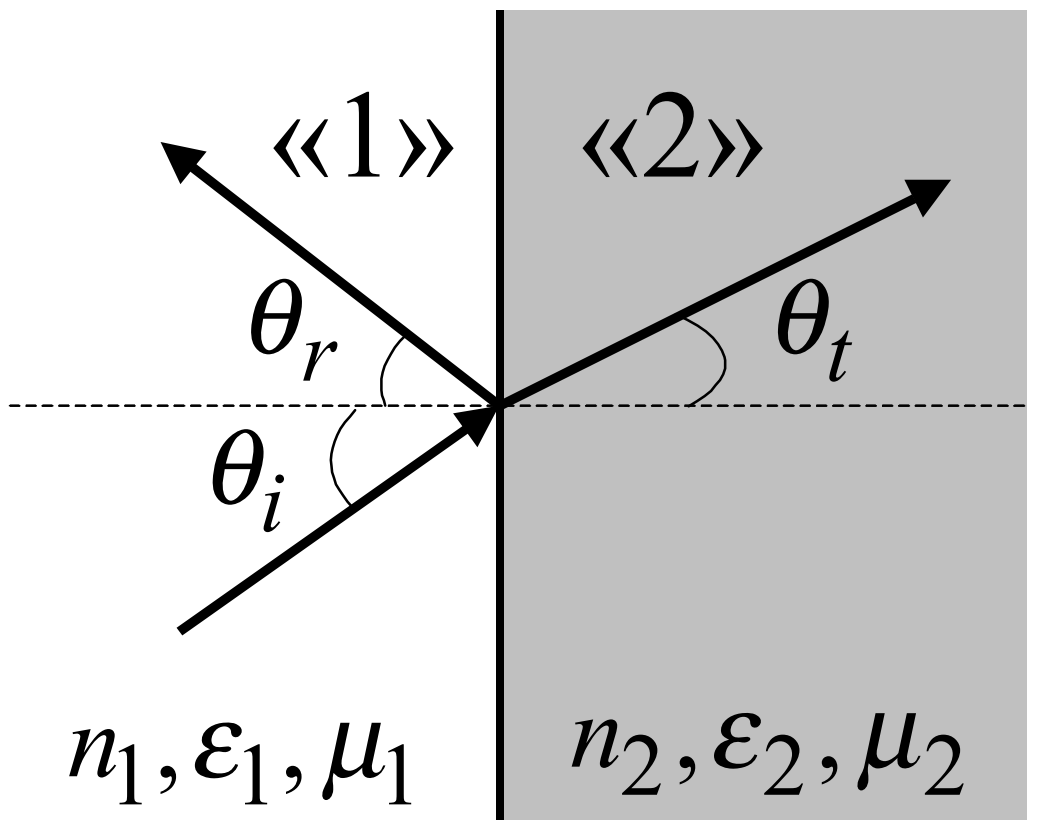

Fig. 2. Geometry of reflection/refraction.

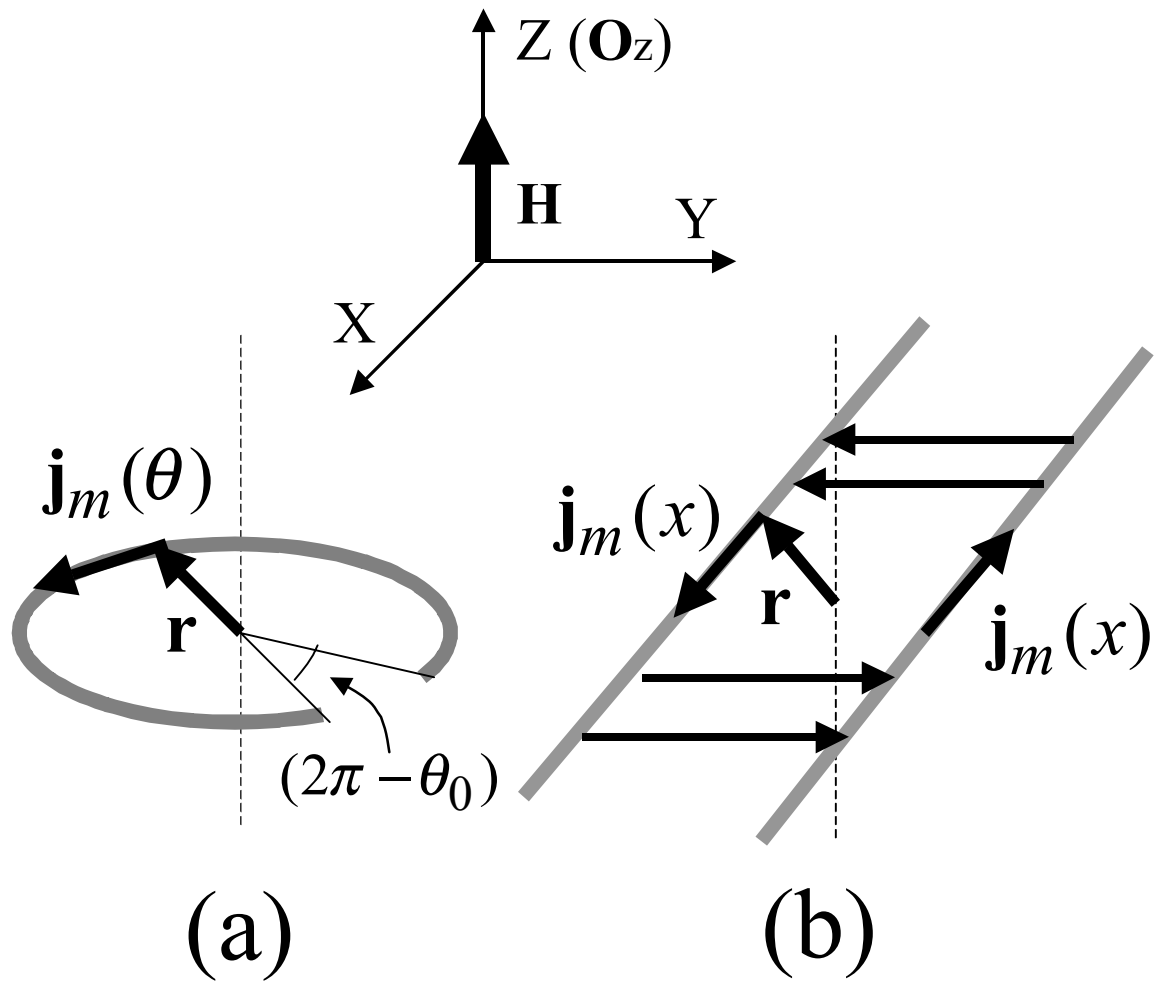

Fig. 3. Geometry of metallic inclusions, principle directions and quantities used.

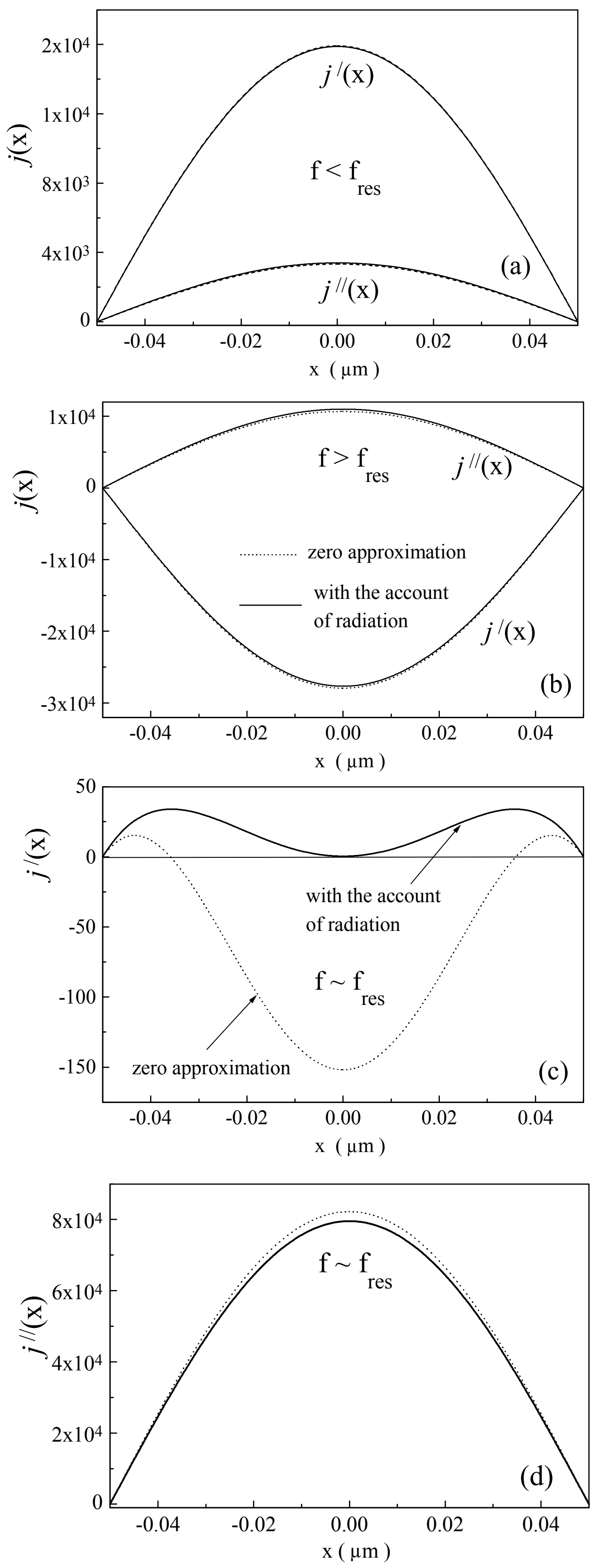

Fig. 4. Typical current distribution (including real $j'$ and imaginary $j''$ parts) along the wire length for different frequencies: (a) $f < f_{res}$, (b) $f > f_{res}$, and (c) and (d) $f \sim f_{res}$.

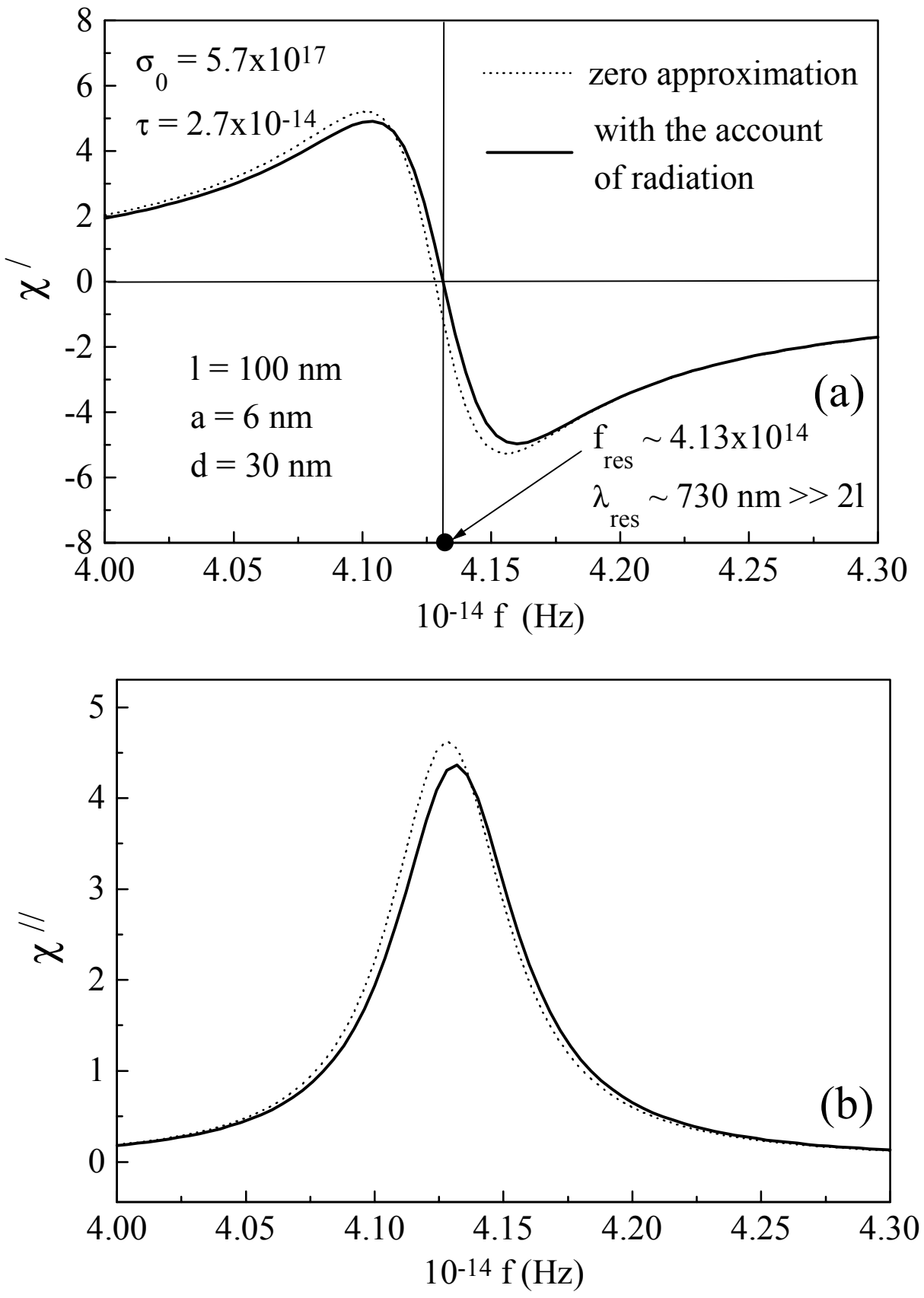

Fig. 5. Magnetic polarisability $\chi_0 = \chi' + i\chi''$ of two-wire contour vs. frequency. The zero (dashed curve) and first (solid curve) approximations are given.

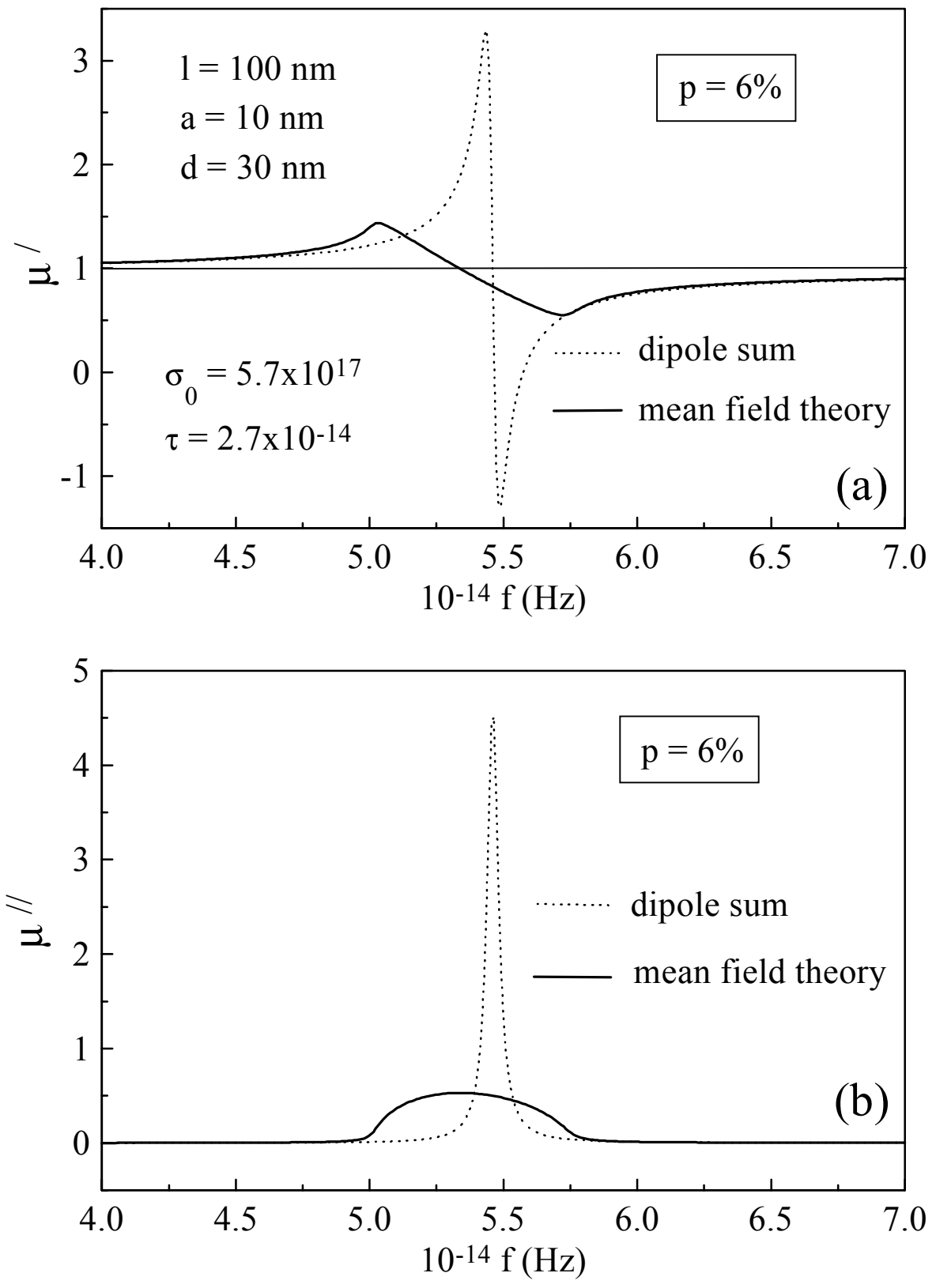

Fig. 6. Effective permeability $\mu_{eff} = \mu' + i\mu''$ of composite containing wire pairs vs. frequency for the volume concentration of 6%, calculated for two cases: independent inclusions (dashed curve) and inclusions in effective medium (solid curve).

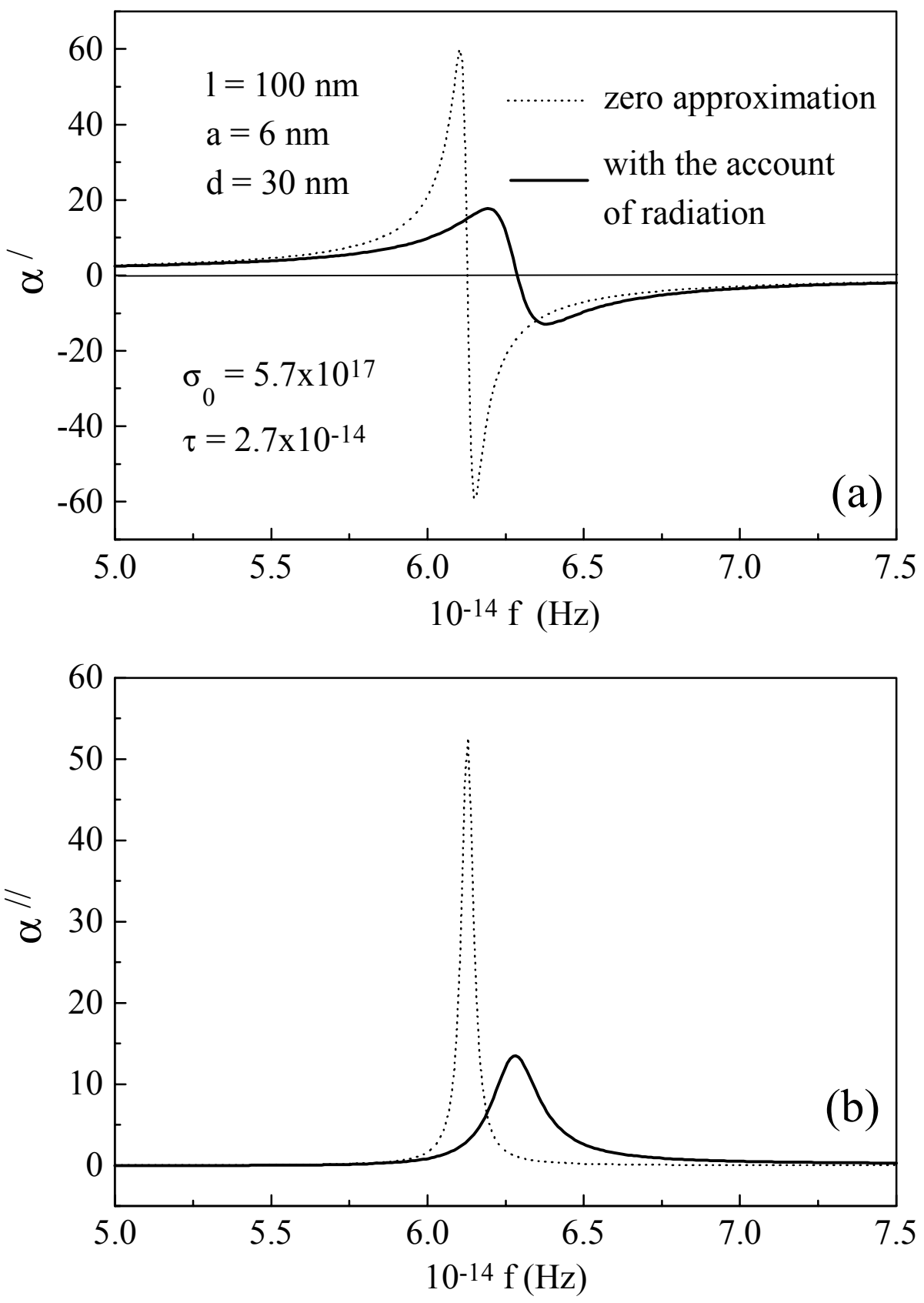

Fig. 7. Electric polarisability $\alpha_0 = \alpha' + i\alpha''$ of two-wire contour vs. frequency. The zero (dashed curve) and first (solid curve) approximations are given.

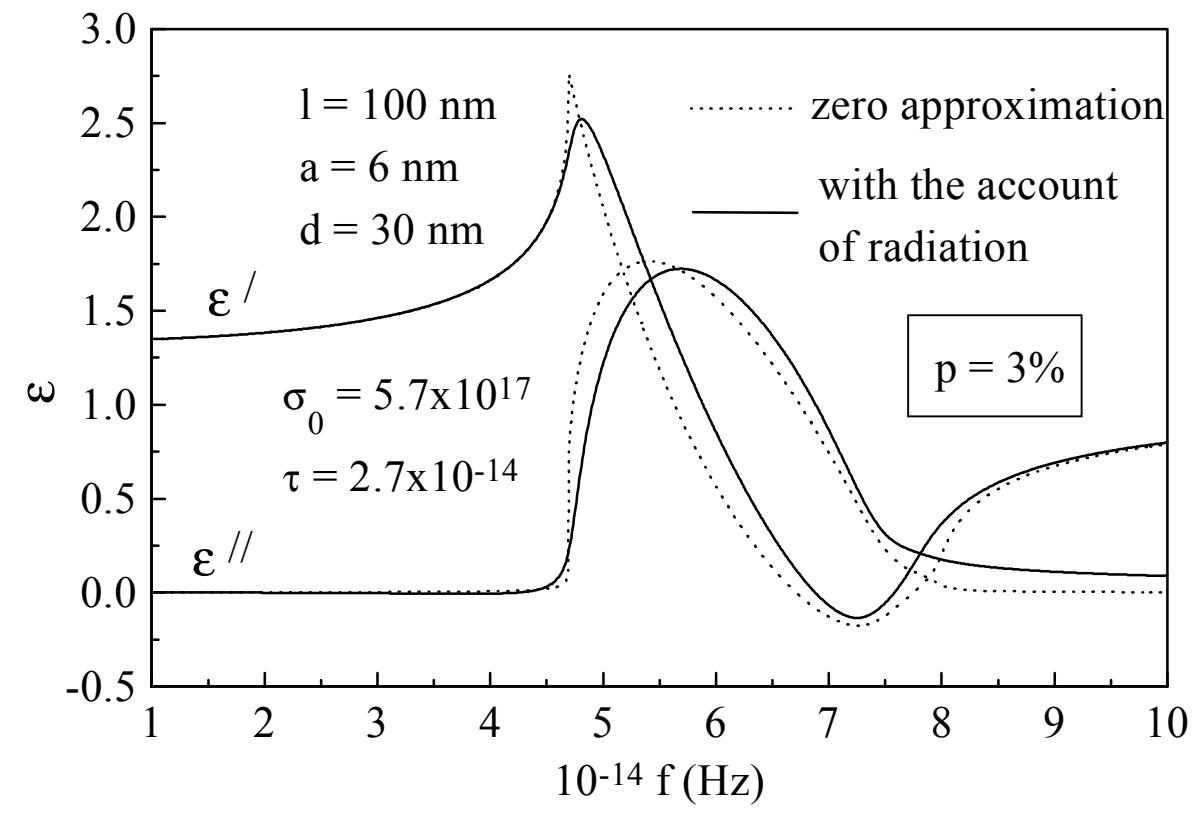

Fig. 8. Effective permittivity of composite containing wire pairs vs. frequency for the volume concentration of 3%. The zero (dashed curve) and first (solid curve) approximations are given.

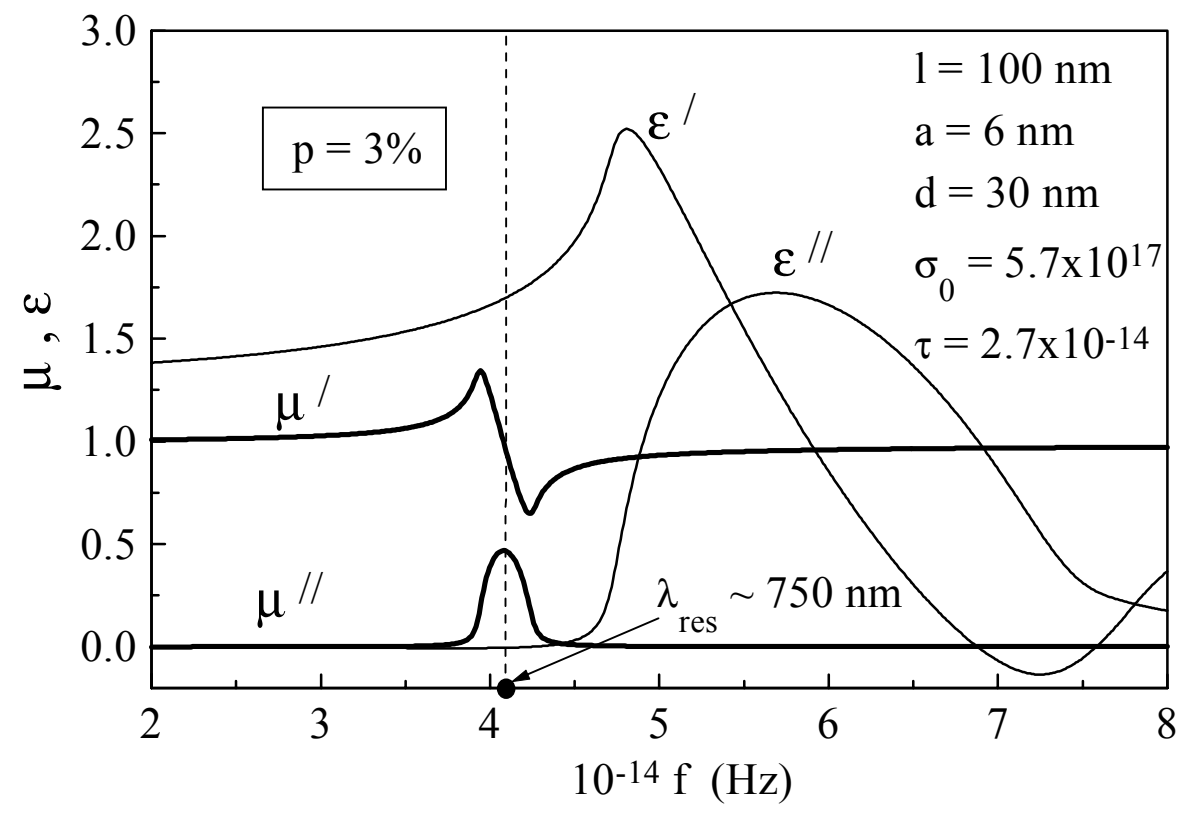

Fig. 9. Effective permeability and permittivity of composite containing wire pairs vs. frequency for the volume concentration of 3%.

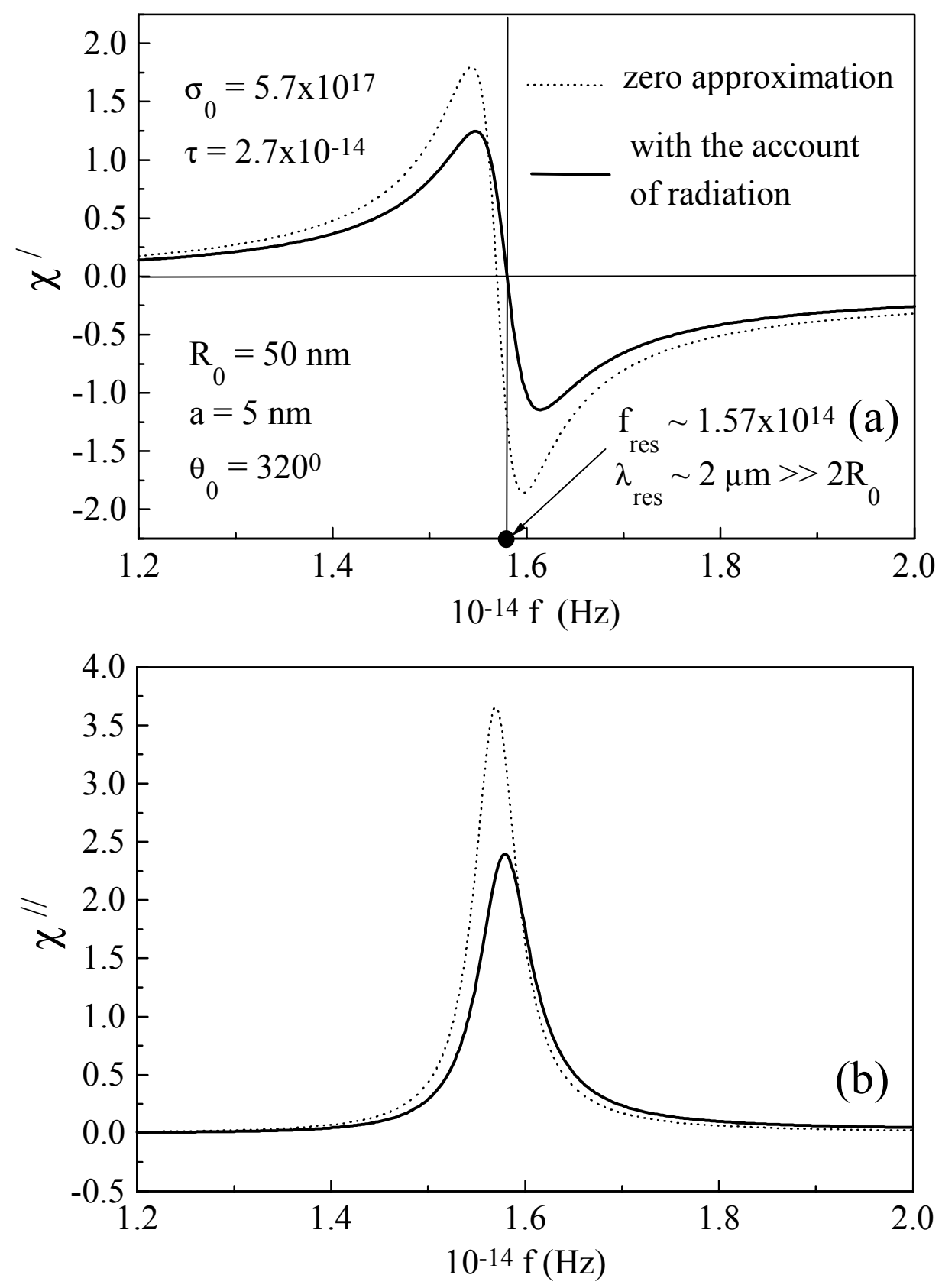

Fig. 10. Magnetic polarisability $\chi_0 = \chi' + i\chi''$ of open- ring contour vs. frequency. The zero (dashed curve) and first (solid curve) approximations are given.

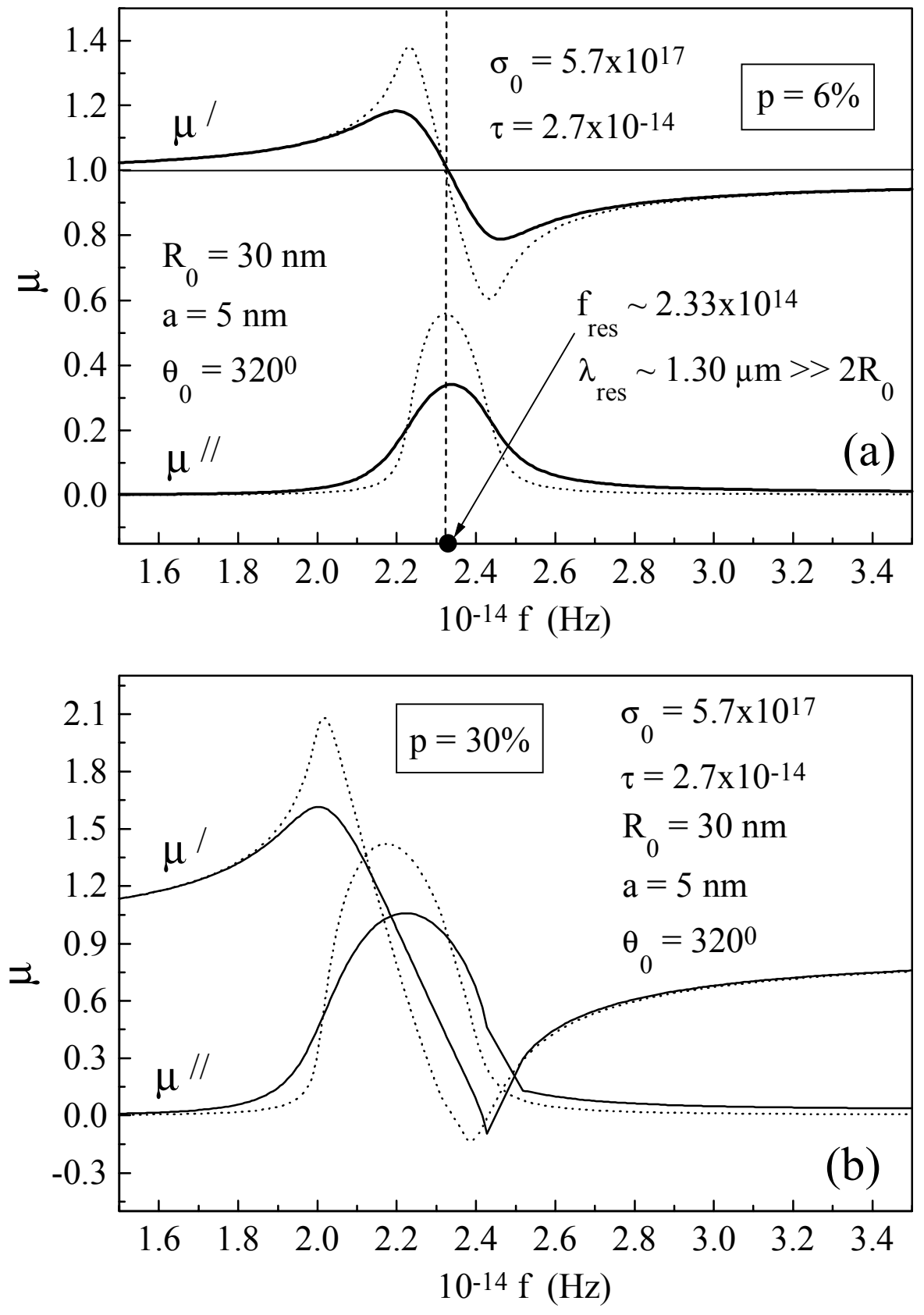

Fig. 11. Effective permeability of composite containing open rings vs. frequency for the volume concentration of 6% (in a) and 30% (in b). The zero (dashed curve) and first (solid curve) approximations are given.

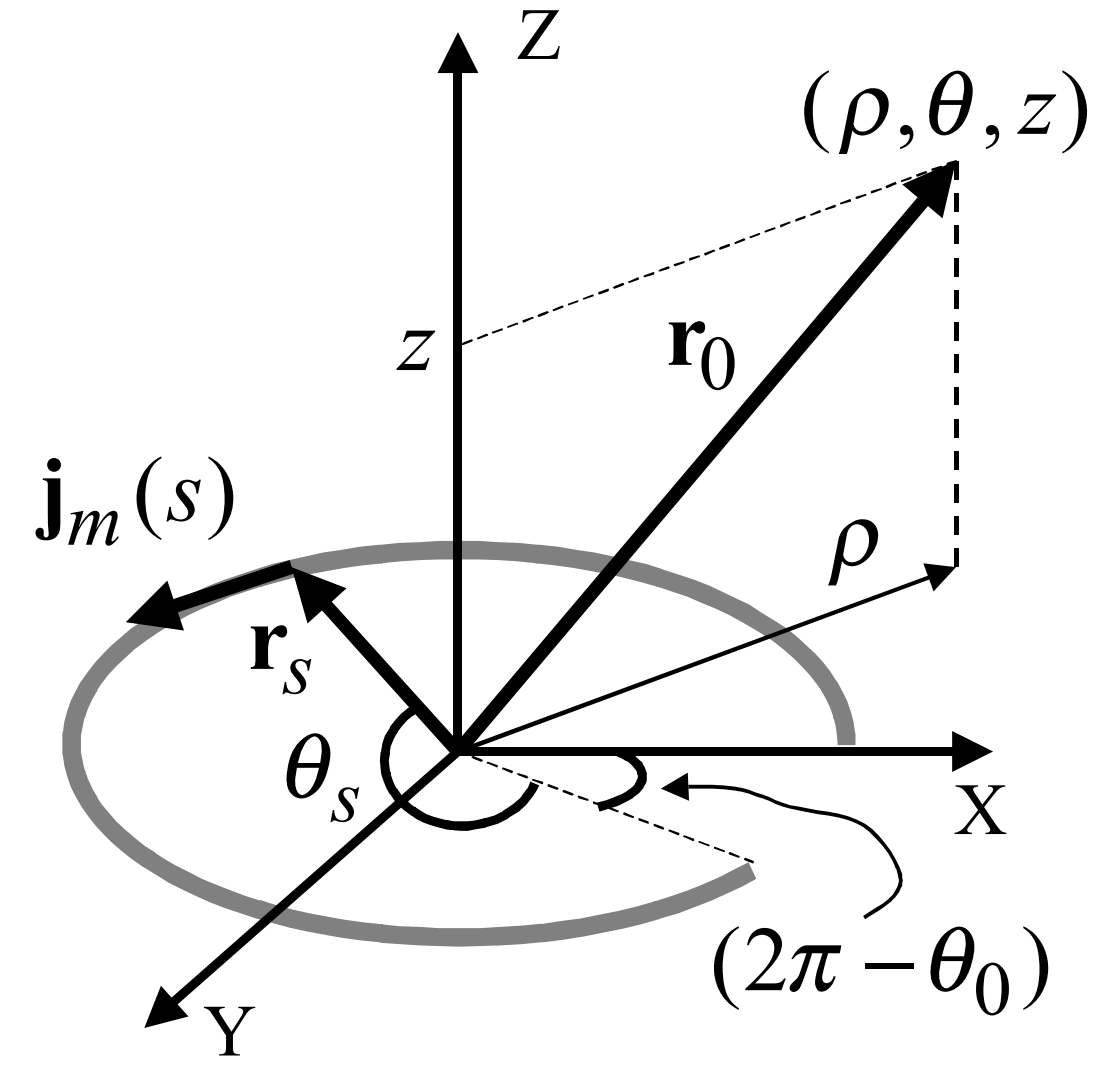

Fig. B1. Principal geometry, directions and quantities used for the calculation of the current distribution along an open-ring.

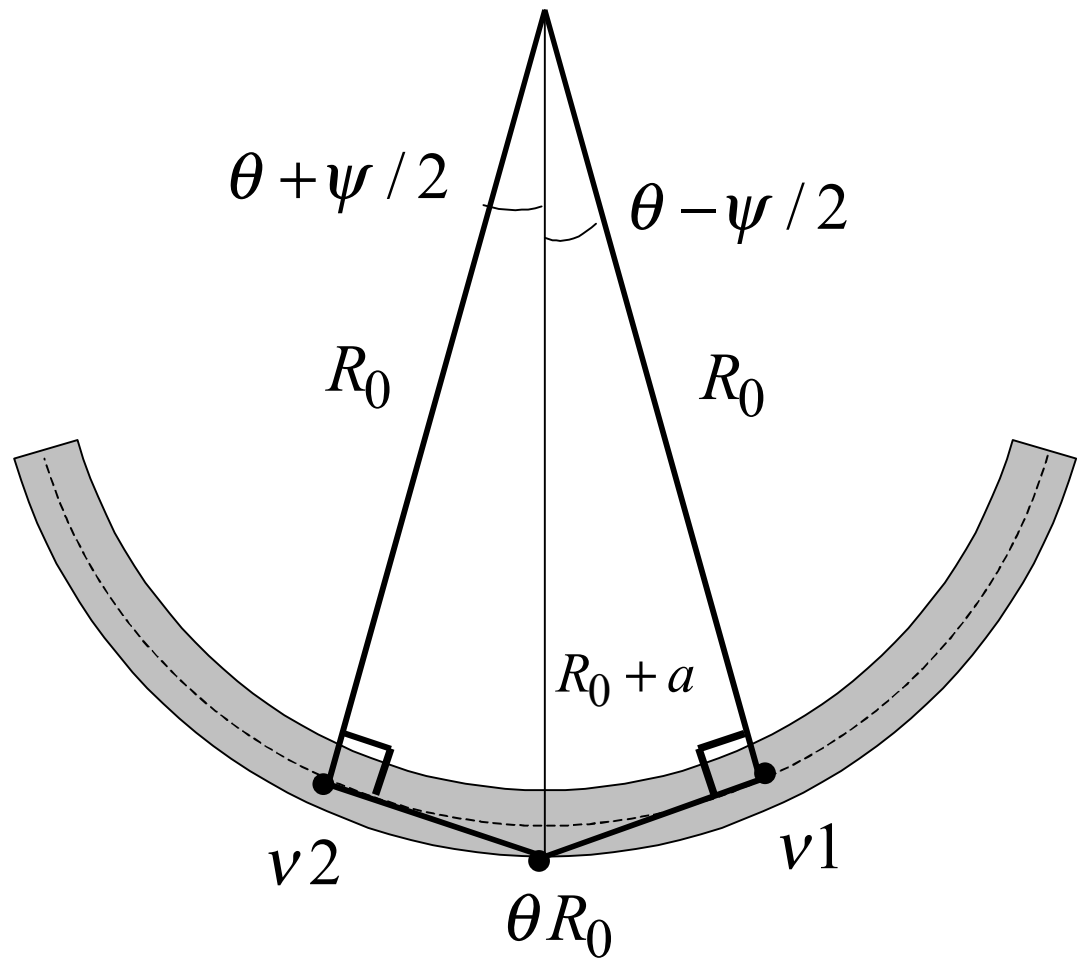

Fig. B2. Principal integration path for equation (B7).